\documentclass[11pt,a4paper]{article}
\usepackage[utf8]{inputenc}
\usepackage[T1]{fontenc}
\usepackage[english]{babel}
\usepackage{lmodern}
\usepackage{calc}
\usepackage{caption}
\usepackage{chngcntr}
\usepackage{enumitem}
\usepackage{setspace}
\usepackage{gensymb}
\usepackage{graphicx} 

\usepackage{listings}

\usepackage{authblk}

\usepackage{layout}
\usepackage{multicol}
\usepackage{multirow}
\usepackage{amsmath,amsfonts,amssymb,amsthm}
\usepackage{mathrsfs}
\usepackage{float}

\usepackage[left=2.5cm,right=2.5cm,top=2.5cm,bottom=2.5cm]{geometry}
\usepackage[final]{pdfpages}
\usepackage{pstricks}
\usepackage{siunitx}

\usepackage{soul}
\usepackage{titlesec}
\usepackage{ulem}
\usepackage{url}

\usepackage{dsfont}
\usepackage{array}

\usepackage{wrapfig}
\usepackage{fancyhdr}
\usepackage[lastpage,user]{zref}

\usepackage{xcolor}
\usepackage{physics}
\usepackage{hyperref}
\usepackage{sectsty}
\definecolor{myblue}{rgb}{0,0.1,0.5}

\usepackage{lineno}
\usepackage[sorting=none]{biblatex} %Imports biblatex package
\hypersetup{
colorlinks=true,
breaklinks=true, 
urlcolor=myblue,
linkcolor=black,
citecolor=myblue,
pdfauthor={Nicolas De Angelis}
}

\usepackage{varwidth}
\DeclareCaptionFormat{myformat}{
  \begin{varwidth}{\linewidth}
    \centering
    #1#2#3
  \end{varwidth}
}

\newcommand{\citefig}[1]{(taken from \cite{#1}, with permission)}  % cite figure taken with permission from a paper
\newcommand{\citefigadapt}[1]{(adapted from \cite{#1}, with permission)}  % cite figure adapted with permission from a paper

\graphicspath{{./Pictures/}}

\begin{document}

\title{Optimizing the light output of a plastic scintillator and SiPM based detector through optical characterization and simulation: A case study for POLAR-2}

% Author list of the POLAR-2 collaboration, usage in main.tex: \input{authors.tex}
\author[a,b]{Nicolas~De~Angelis\footnote{Corresponding author. \newline \href{mailto:nicolas.deangelis@inaf.it}{nicolas.deangelis@inaf.it}}} % 0000-0002-2498-0213
\author[a]{Franck~Cadoux}  % franck.cadoux@unige.ch
\author[a]{Coralie~Husi}  % coralie.husi@unige.ch
\author[a,c]{Merlin~Kole}  % merlin.kole@unige.ch  % 0000-0003-0441-4959
\author[d]{Sławomir~Mianowski}  % slawomir.mianowski@ncbj.gov.pl % 0000-0003-2514-6156
%\author[d]{J.M.~Burgess}
%\author[d]{J.~Greiner}
%\author[a]{J.~Hulsman}  % johannes.hulsman@unige.ch
%\author[b]{H.C.~Li}
%\author[g]{J.~Mietelski}
%\author[c]{A.~Pollo}
%\author[b]{N.~Produit}
%\author[b]{D.~Rybka}  % dominik.rybka@ncbj.gov.pl
%\author[a]{J.~Stauffer}
%\author[e]{J.C.~Sun}
%\author[e]{B.B.~Wu}
%\author[a]{X.~Wu}
%\author[e,f]{S.N.~Zhang}
\affil[a]{DPNC, University of Geneva, 24 Quai Ernest-Ansermet, CH-1205 Geneva, Switzerland}
%\affil[b]{Geneva Observatory, ISDC, University of Geneva, 16, Chemin d’Ecogia, CH-1290 Versoix, Switzerland}
\affil[b]{INAF-IAPS, via del Fosso del Cavaliere 100, 00133 Rome, Italy}
\affil[c]{Space Science Center, University of New Hampshire, Durham, NH 03824, USA}
\affil[d]{National Centre for Nuclear Research, ul. A. Soltana 7, 05-400 Otwock, Swierk, Poland}
%\affil[e]{United Kingdom Atomic Energy Authority, Culham Science Centre, Abingdon OX14 3DB, UK}
%\affil[d]{Max-Planck-Institut fur extraterrestrische Physik, Giessenbachstrasse 1, D-85748 Garching, Germany}
%\affil[e]{Key Laboratory of Particle Astrophysics, Institute of High Energy Physics, Chinese Academy of Sciences, Beijing 100049, China}
%\affil[f]{University of Chinese Academy of Sciences, Beijing 100049, China}

\renewcommand{\thefootnote}{\arabic{footnote}}

\date{\today}

\maketitle

%old title: Surface roughness characterization of diamond-milled plastic scintillators
%since we only measured a few diamond milled scintillators (others were as cast, not polished), this paper will discuss all the optical sims, so new title is more general about optical simulations for POLAR-2 polarimeter modules

\vspace*{-0.5cm}
\section*{Abstract}

The combination of plastic scintillators with Silicon Photo-Multipliers (SiPMs) is widely used for detecting radiation in high-energy astrophysics, particle physics, neutrino physics, or medical physics. An example of application for this kind of detectors are Compton polarimeters such as POLAR-2 or LEAP, for which a low-Z material is needed for the Compton effect to be dominant down to as low energy as possible. Such detectors aim to measure low energy Compton depositions which produce small amounts of optical light, and for which optimizing the instrumental optical properties consequently imperative.\\

The light collection efficiency of such a device was studied with a focus on the POLAR-2 Gamma-Ray Burst polarimeter. POLAR-2 consists of a segmented array of 6400 elongated plastic scintillators divided into 100 modules, all read out by SiPMs. The conversion of incoming $\gamma$-rays into readable signal goes through the production and collection of optical light, which was to be optimized both through measurements and simulations. The optical elements of the POLAR-2 polarimeter prototype module were optically characterized and an optical simulation based on Geant4 was developed to fully model its optical performances. The results from simulations were used to optimize the design and finally to verify its performance. The study resulted in a detector capable of measuring energy depositions of several keV. In addition an important finding of this work is the impact of the plastic scintillator surface roughness on the light collection. It was found that a plastic scintillator with a higher scintillation efficiency but made of a softer material, hence with a rougher surface, was not necessarily the best option to optimize the light collection. Furthermore, in order to optimize the optical crosstalk between different channels, a production technique for very thin ($\sim$150$~\mu$m) and reusable silicone-based optical coupling pads was developed. This method can be adapted to produce either standalone pads or to directly mold a layer on any SiPM to be later coupled to the scintillators.\\

After an introductory discussion describing the need of a large scale GRB polarimeter like POLAR-2, the optical design and characterization of the polarimeter modules which composes its sensitive part are described. The Geant4-based optical simulations of the POLAR-2 modules and the impact of the optical properties of its various elements on the light collection efficiency of the instrument are later presented. The work is finally summarized and an outlook is given on the potential applications of the POLAR-2 optical characterization and simulation work to other experiments employing similar elements.\\

\noindent\textit{Keywords:} Detector modelling and simulations I (interaction of radiation with matter, interaction of photons with matter, interaction of hadrons with matter, etc); Scintillators, scintillation and light emission processes (solid, gas and liquid scintillators); Polarimeters; Simulation methods and programs

\tableofcontents

\newpage
\section{GRB polarimetry with POLAR-2 and the need for a detailed optical characterization of its polarimeter}

Gamma-Ray Bursts (GRBs) are very extreme phenomena of great interest in modern astrophysics as they provide a unique place to probe extreme environments, both on a physical and astrophysical point of view, and make a perfect candidate for multi-messenger and multi-band observations. Although discovered almost 70 years ago, these extragalactic transient phenomena remain a considerable source of unresolved questions, notably concerning the responsible emissions mechanisms as well as the jet and magnetic field structures of these cosmic objects. Measuring the polarization of their prompt $\gamma$-ray emission could help answering many of these questions, justifying the need for a dedicated GRB polarimeter. After the success of POLAR \cite{POLAR_detector, POLAR_catalog} and based on its legacy, a larger scale Compton polarimeter has been developed to be placed on the China Space Station in 2027.\\

Compton scattering being the dominant photon interaction process in the prompt emission energy range (tens to hundreds of keV), it is used as a way to detect polarization. In order for the Compton effect to be dominant down to as low energies as possible\footnote{Since the spectrum of a GRB typically follows a power-law at a few tens of keV, lowering the energy threshold of the instruments greatly improves its sensitivity as the number of detected counts for a given source dramatically increases.}, a low-Z material has to be employed. Plastic scintillators are therefore used to convert the deposited $\gamma$-ray energy into optical photons, which are in turn read out by Silicon PhotoMultipliers (SiPMs). These type of scintillators having a very low scintillation efficiency\footnote{Meaning that a smaller amount of optical light is produced per unit of deposited energy compared to an inorganic scintillator.}, optically characterizing and simulating such a detector is imperative in order to optimize the light collection efficiency and to be able to convert low-energy depositions (down to a few keVs) into a measurable signal. This is a crucial point to improve the sensitivity of any Compton polarimeter, or more generally any plastic scintillator-based instrument.

\begin{figure}[H]
\centering
\includegraphics[width=0.9\textwidth]{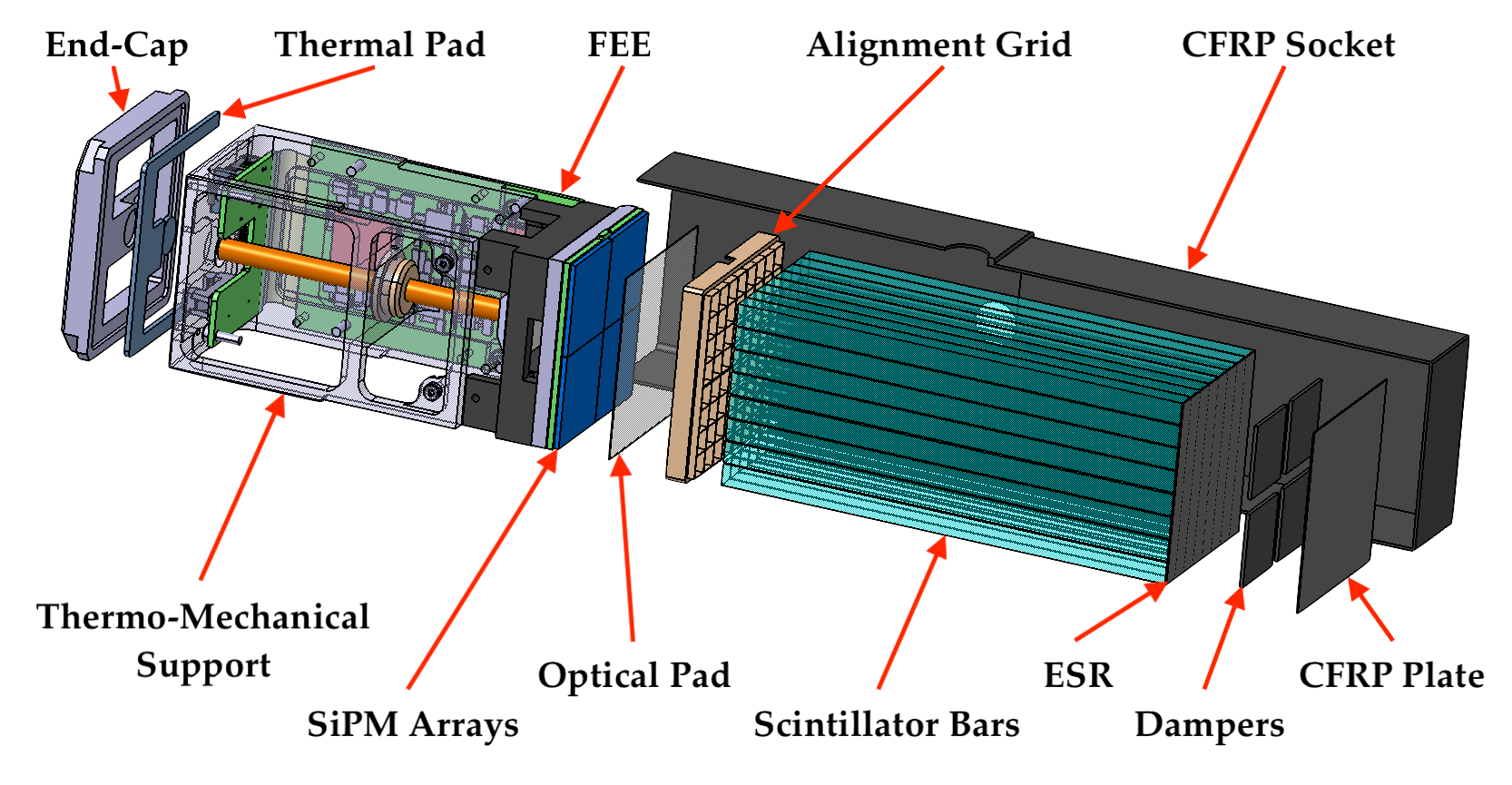}
  \caption{Exploded view of the POLAR-2 polarimeter module CAD model \citefig{NDA_thesis}}
  \label{fig:polar-2_module_design}
\end{figure}

Figure \ref{fig:polar-2_module_design} shows a CAD exploded view of the POLAR-2 polarimeter module. The module can be divided into two parts: \textit{(i)} the sensitive part, of interest in this work, which consists of an array of plastic scintillators read out by SiPM arrays; and \textit{(ii)} the readout electronics and its thermo-mechanical frame. As described in the next section, several improvements have been brought to the design of the first part (compared to POLAR \cite{POLAR_detector}) in order to optimize the light collection efficiency and therefore improve the sensitivity of the instrument. %Indeed, one should ensure an optimal optical behavior of the detector through both detailed characterizations and simulations as the energy deposited by the incoming $\gamma$-rays is converted into optical light by the scintillators which in turn has to be detected.
In particular, these upgrades allowed to increase the instrument sensitivity to typical GRB spectra by an order of magnitude while the number of channels is only quadrupled, and to improve the modulation factor\footnote{The modulation factor, denoted here as $\mu_{100}$, corresponds to the relative amplitude of the modulation observed in the azimuthal scattering angle distribution as measured for a fully polarized source. It gives an indication on the sensitivity to polarization.} \cite{ESRF_paper, NDA_thesis}.\\

After a comprehensive discussion of the optical characterization work conducted for optimizing the polarimeter module design, the Geant4-based optical simulations of the module are presented. The physical performances of the detector obtained both through simulations and calibration measurements are finally compared.

\newpage
\section{Optical design and characterization of the polarimeter module}\label{sec:optical_design_POLAR-2}

\begin{figure}[H]
\centering
  \includegraphics[height=0.4\textwidth, angle=270]{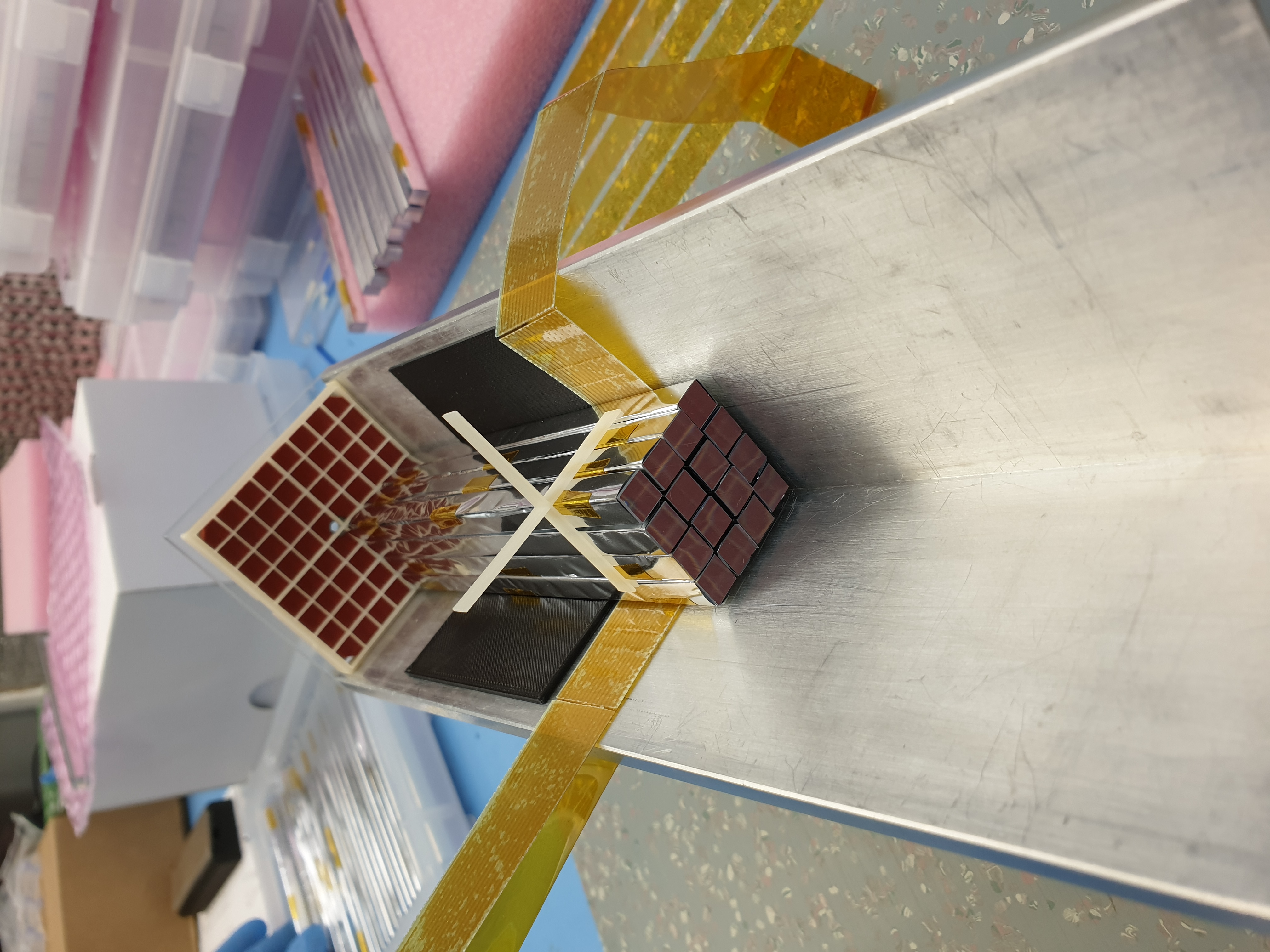}\hspace*{0.1cm}\includegraphics[height=0.4\textwidth, angle=270]{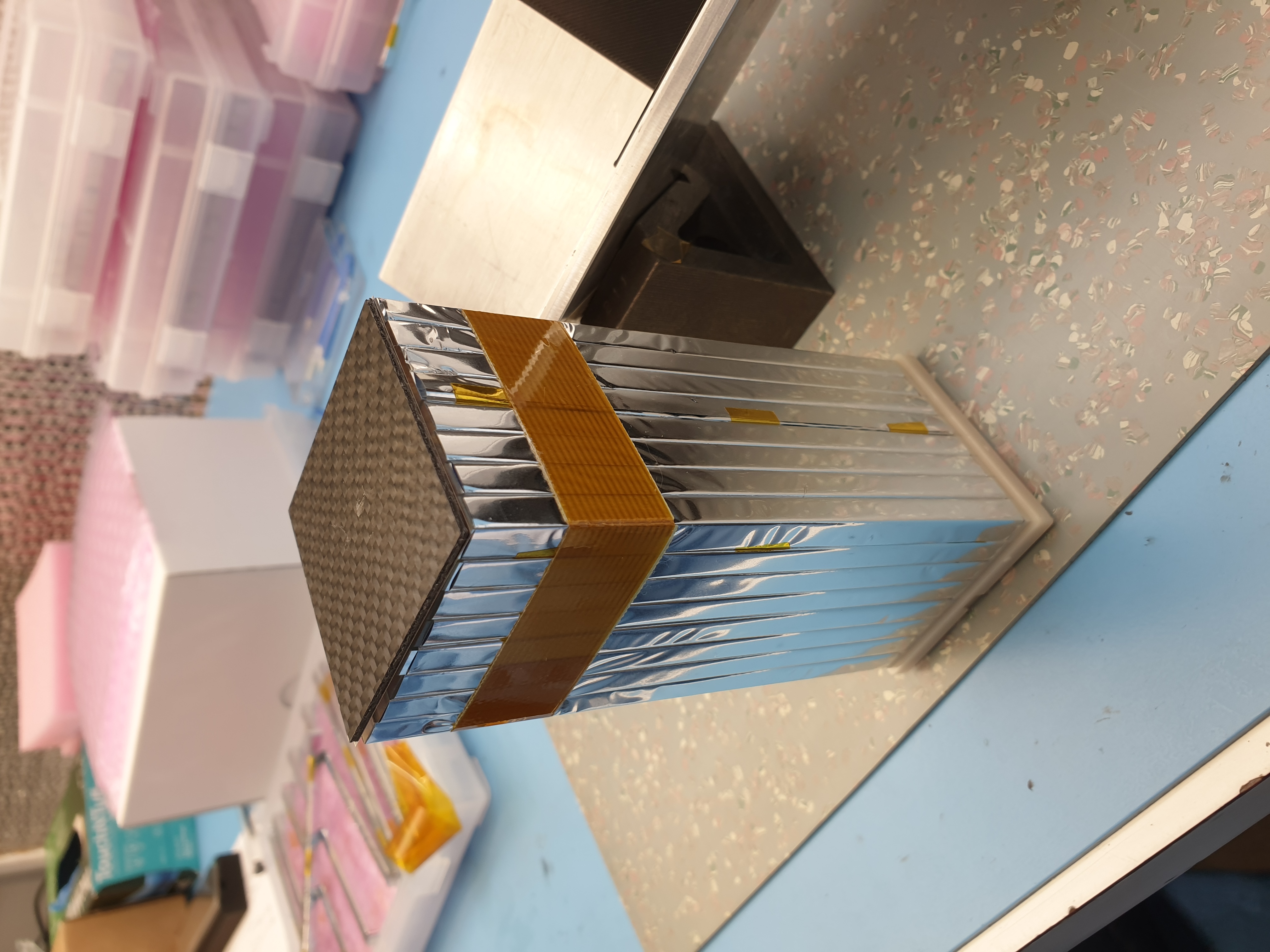}
  \caption{\textbf{Left:} A quarter of a target being assembled in the plastic alignment grid. \textbf{Right:} A fully assembled target ready to be integrated into a polarimeter module. \citefig{NDA_thesis}}
  \label{fig:target_assembly}
\end{figure}

\begin{figure}[H]
\centering
  \includegraphics[height=0.28\textwidth]{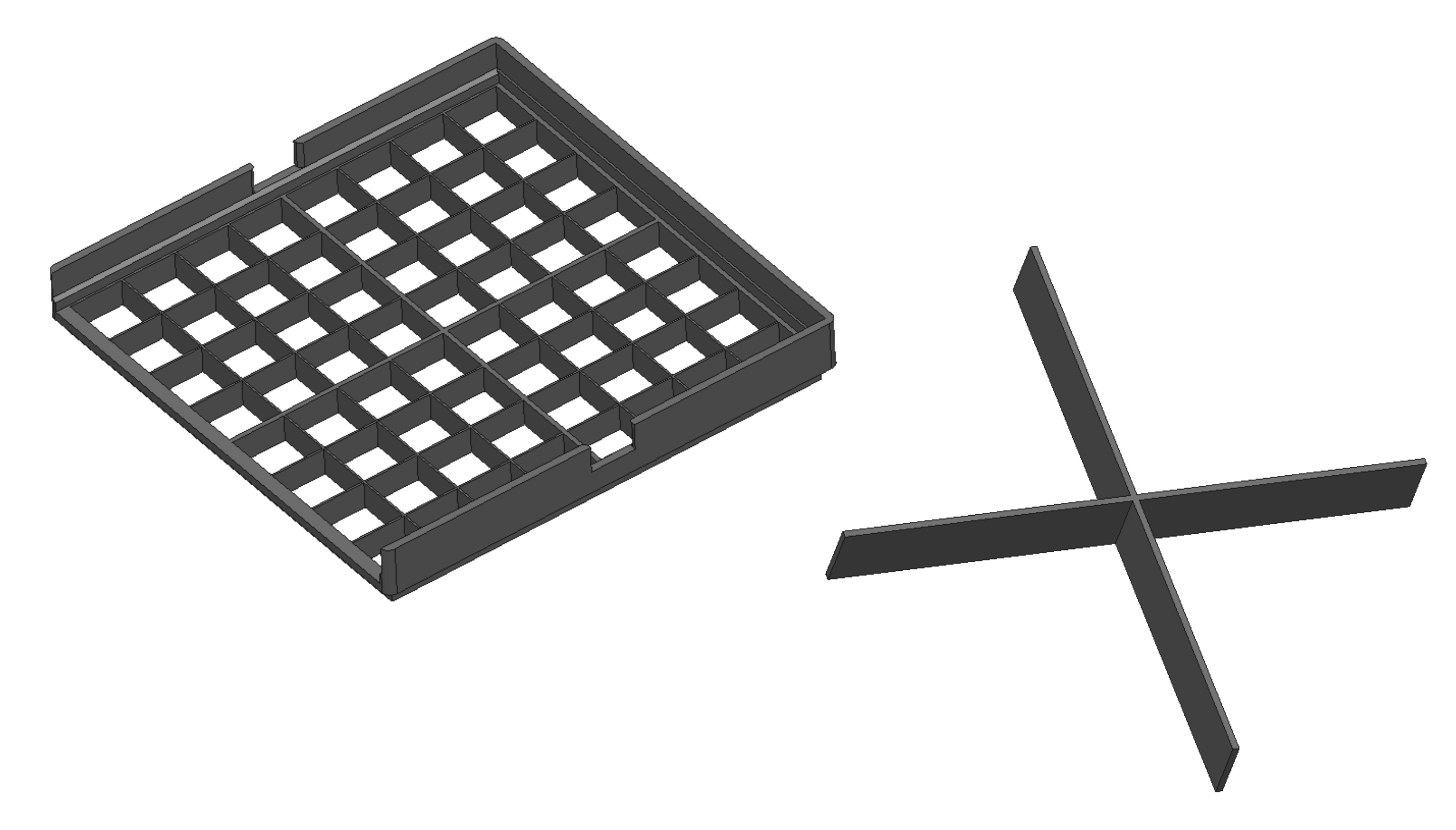}\includegraphics[height=0.28\textwidth]{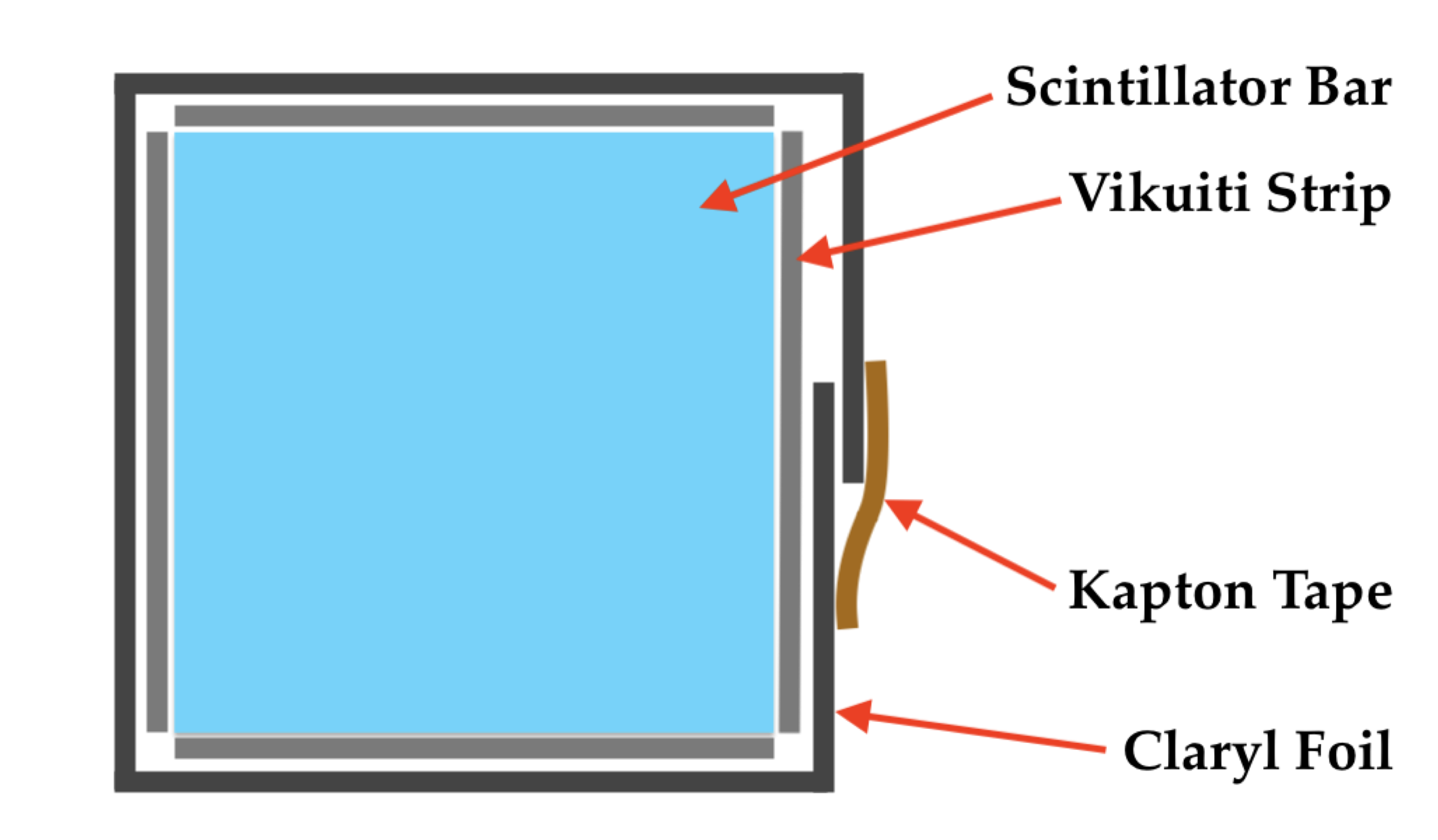}
  \caption{\textbf{Left:} CAD design of the plastic alignment grid and cross \citefig{NDA_thesis}. \textbf{Right:} Schematic view of the disposition of the different wrapping layers around a scintillator bar \citefig{NDA_thesis}.} 
  \label{fig:grid_design_wrapping}
\end{figure}

As in POLAR \cite{POLAR_detector}, the POLAR-2 polarimeter module consists of an 8$\times$8 array of elongated plastic scintillator bars read out by light sensors. The incoming $\gamma$-ray is depositing energy into the scintillators which is converted into optical light and collected by the optical sensors. However, the design of this module has been considerably improved compared to POLAR in order to optimize the light collection as well as the physical performances of the polarimeter. As shown in Figure \ref{fig:polar-2_module_design}, the employed light sensors are no longer Multi-Anode Photo-Multiplier Tubes (MA-PMTs), but Silicon Photo-Multipliers (SiPMs), which have a twice higher Photo-Detection Efficiency (PDE) \cite{Hamamatsu_datasheet_pmt, Hamamatsu_datasheet_s13series} and show many other advantages (e.g. compactness, lower bias voltage, insensitivity to magnetic fields). Another benefit from the use of SiPMs is that their entrance coating is much thinner than the borosilicate entrance window of MA-PMTs (about 0.1~mm against 1.5~mm), which combined with the development of a new production technique \cite{Module_assembly_TN, Coralie_optpad} for the optical coupling pads between the scintillators and the sensors (allowing to reduce their thickness from 0.7~mm down to 150~$\mu$m, see Section \ref{subsec:opt_pad_meas}), greatly reduced the optical crosstalk between channels. The optical crosstalk was also improved thanks to the development of a new wrapping technique \cite{Module_assembly_TN}, which allows to individually wrap each scintillator bar, as shown in Figures \ref{fig:target_assembly} and \ref{fig:grid_design_wrapping}.

The plastic grids used to hold and align the 64 wrapped scintillators were redesigned (see Figure \ref{fig:grid_design_wrapping}), allowing to increase the cross-section of the bars from 5.8$\times$5.8~mm$^2$ to 5.9$\times$5.9~mm$^2$ and therefore reducing the dead-space in the instrument and increasing its sensitive volume\footnote{One should note that the length of the scintillator bars was also reduced from 176~mm to 125~mm to improve the singal-to-noise ratio \cite{scint_length_TN}}. Furthermore, thanks to the design optimization of the plastic alignment grids and the high precision of the 3D-printing process used to manufacture them, allowing to print parts as thin as 200~$\mu$m, the tapering that was present on both side\footnote{While in POLAR a mechanical grid was present on both ends of the scintillators, the thinner upgraded alignment grid for POLAR-2 is only needed on the SiPM side, while the other side just requires the use of a middle plastic cross to match the extra-spacing between the 4 SiPM arrays used to readout the target.} of the scintillator bars of POLAR  for mechanical reasons \cite{POLAR_detector} have been removed for POLAR-2. This allowed to increase the contact surface between the scintillator bar and the SiPM by about 40\% (from 5.0$\times$5.0~mm$^2$ to 5.9$\times$5.9~mm$^2$).

In this section, the focus is given to the optical characterization of various optical elements of the POLAR-2 polarimeter module, while the next section discusses their implication on the module design both through optical simulations and calibration measurements.

\subsection{Reflective foils}\label{subsec:ESR_characterization}

The role of the reflective foils is to keep as many optical photons as possible inside the scintillators until they reach the SiPMs where they get absorbed through the photo-electric effect and converted into a readable signal. Maximizing the reflectivity in the 400-500~nm wavelength interval, corresponding to emission wavelengths of the plastic scintillators, is therefore decisive on the overall optical efficiency of the module. As a consequence, the reflectivity and transmissivity of several kinds of highly reflective films were measured \cite{ESR_TN} at the Optical Quality Control Lab of the CERN Thin Film \& Glass service using a Perkin Elmer Lambda650 spectrometer \cite{Lambda_spec_datasheet}. This spectrometer allows to measure the specular reflectivity with the so-called Universal Reflectance Accessory (URA), as well as the total reflectivity and its diffuse component using a 150~mm diameter integration sphere \cite{NDA_thesis}. Using this latter setup, the transmittance of the foil can also be characterized.\\

\begin{figure}[H]
\centering
 \includegraphics[width=.6\textwidth]{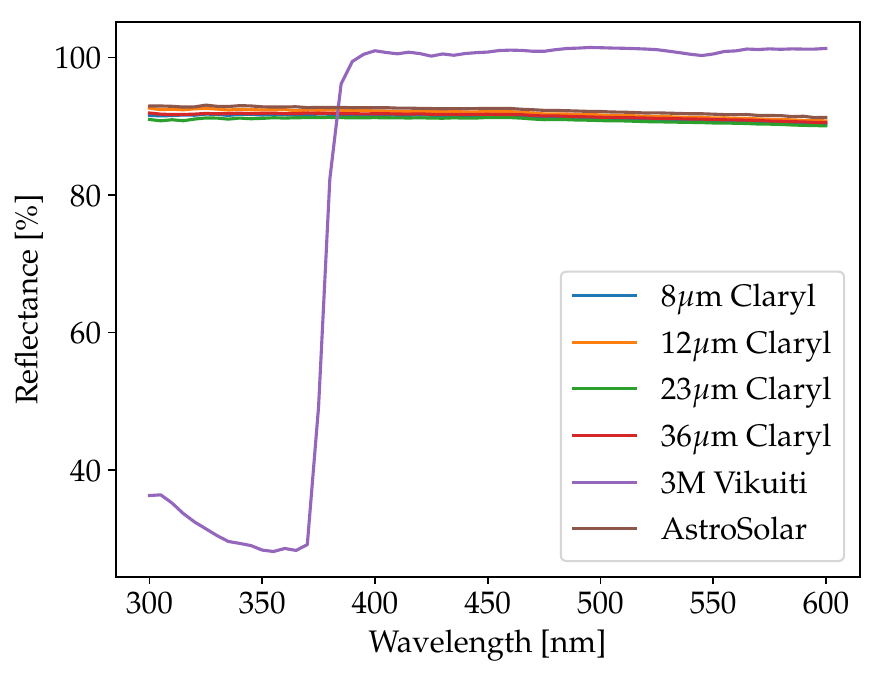}
 \caption{Total reflectance spectrum of all the Claryl foils compared to that of Vikuiti and AstroSolar \citefig{NDA_thesis}}
 \label{fig:tot_refl}
\end{figure}

Several samples of highly reflective foils were investigated as candidate for the wrapping of scintillators. The first tested foil is the Vikuiti from 3M\footnote{\url{www.3m.com}}, a 65~$\mu$m thick Enhanced Specular Reflector film (ESR) which was previously used in POLAR. The Claryl film from Toray\footnote{\url{www.toray.eu/eu/}}, an aluminized PET layer available in several thicknesses from 8 to 36~$\mu$m, was also tested as a reflector. Finally, the Astrosolar\textregistered{} foil from Baader Planetarium\footnote{\texttt{\url{www.baader-planetarium.fr}}}, used for optical Solar observations, was also considered. The total reflectance of all of these reflective foil was measured in the 300-600~nm range and is plotted in Figure \ref{fig:tot_refl}.

Both the Astrosolar and Claryl samples, for different thicknesses of PET substrate for the latter, show a similar total reflectance. The reflectance is between 90 and 94\% with a very small decrease with increasing wavelength, typical of Aluminum reflectors. The PET layer thickness does not have a significant effect on the reflectance, since it is the Aluminum coating on top of the substrate that gives its reflectivity to the sample. The measured Claryl reflectance matches the expected reflectance from Aluminum \cite{Hass:61}. For the Vikuiti sample, the behavior is completely different: above 380~nm, the total reflectance is compatible with 100\%, and the precision of the spectrometer does not actually allow to disentangle the sample reflectance with a perfect reflection. This difference is due to the fact that the Vikuiti foils are not coated with Aluminum as the other samples, but are instead entirely composed of plastic. This cutoff being below 400~nm it is not relevant for POLAR-2 as the emission range of the plastic scintillators used for the polarimeter is 400-500~nm (see Figure \ref{fig:scintillator_emission_spec_sim}). Details on the specular and diffuse components of these reflectors are provided in Appendix \ref{sec:appendix_ESR}, together with a comparison of the reflectivity between the two sides of the foil. \\

\begin{figure}[H]
\centering
 \includegraphics[width=\textwidth]{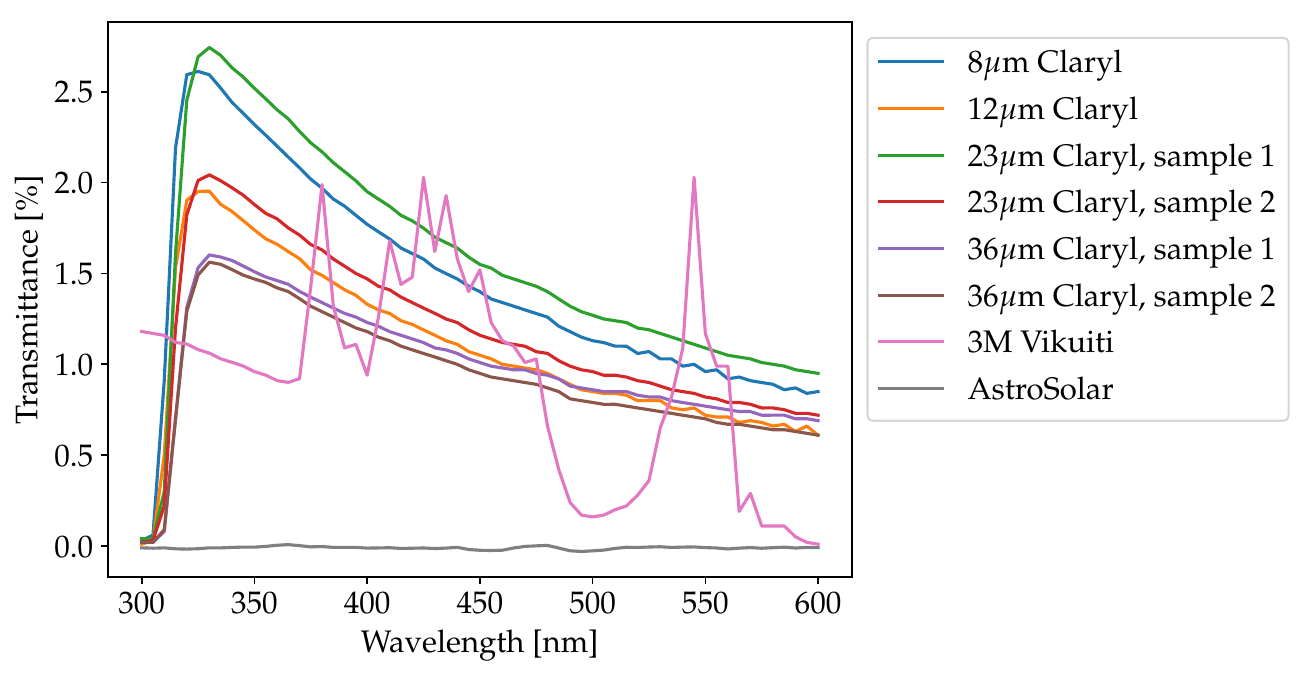}
 \caption{Transmittance spectrum measured for Vikuiti, Claryl, and Astrosolar samples \citefig{NDA_thesis}.}
 \label{fig:trans_ESR}
\end{figure}

In addition to keeping the light inside the scintillators to maximize light collection, the role of the reflective foils is also to prevent photons from going to the neighbor channels, causing optical crosstalk. The transmittance of the considered reflectors is in consequence a key parameter to be characterized as well. The measured spectral transmittance for different foils are plotted in Figure \ref{fig:trans_ESR}. The averaged transmittance in the 400-500~nm range is around 1-2\% for both Claryl and Vikuiti, while the Astrosolar foil has a transmittance smaller than the resolution of the spectrometer and seems therefore consistent with no transmission. The result for the Astrosolar foil was to be expected since it has been design as a filter with optical density of 5 to observe the Sun, meaning that only $10^{-5}$ of the photons will be transmitted though the foil.

Even though the transmittances for the Claryl and Vikuiti are in the same order of magnitude, we can observe a more complex structure in the Vikuiti curve. This is due to the higher complexity of the Vikuiti foil, which is made of numerous layers of plastics playing with interferences to reflect back almost all the photons. No complex structure is observed for the Claryl since it just consists of a PET transparent layer coated on one side with Aluminum.\\

Since the Vikuiti has the best reflectivity performances of all the tested reflective foils, it has been selected as the primary reflector for wrapping the POLAR-2 scintillator bars. Nevertheless, because of its mechanical rigidity, it cannot be folded around the bar to cover the 4 sides. Four 5.9$\times$122~mm strips of the 65~$\mu$m thick Vikuiti are therefore used to cover the long sides of the bar. A single 25$\times$122~mm foil of 8~$\mu$m thick Claryl is then used to wrap the scintillator and the 4 Vikuiti strips and hold them together, as depicted in Figure \ref{fig:grid_design_wrapping}. Each bar can thus be wrapped individually, with the Claryl preventing photons to escape from small gaps between Vikuiti strips in the corners of the bar. Since two layers of Vikuiti and two layers\footnote{See Appendix \ref{sec:appendix_ESR} for a transmittance measurement of a double layer of Claryl.} of Claryl are separating each bars, almost no crosstalk\footnote{Assuming a transmittance of 1\% for both foils, the expected crosstalk over the wrapped fraction of the bar with 4 layers separating neighbor bars is $0.01^4=10^{-8}$.} is expected from the 122~mm (out of the total 125~mm of the bar length) covered by reflective foils. The remaining 3~mm are covered by the plastic alignment grid used to assemble all the bars together. The optical properties of the grid material were therefore studied.

\subsection{Plastic alignment grids}\label{subsec:plastic_grid}

As it covers 2.4\% of the height of the scintillators\footnote{The grid is 3~mm thick, and the scintillator bars are 125~mm long. See Figure \ref{fig:grid_design_wrapping} for the grid design.}, characterizing the transmittance of the plastic composing the alignment plastic grids is important for understanding all the contributions to the inter-channels optical crosstalk. 20$\times$20~mm$^2$ plastic plates of different thicknesses made with the same material as the alignment grid were therefore printed at the CERN Polymer Lab in order to measure the transmittance of the material. The accuracy on the sample thickness achieved with the 3D printer is better than 50~$\mu$m. The same setup as used in the previous section to characterize the reflectors is used to measure the transmittance of the plastic samples.

\begin{figure}[H]
\centering
  \includegraphics[width=\textwidth]{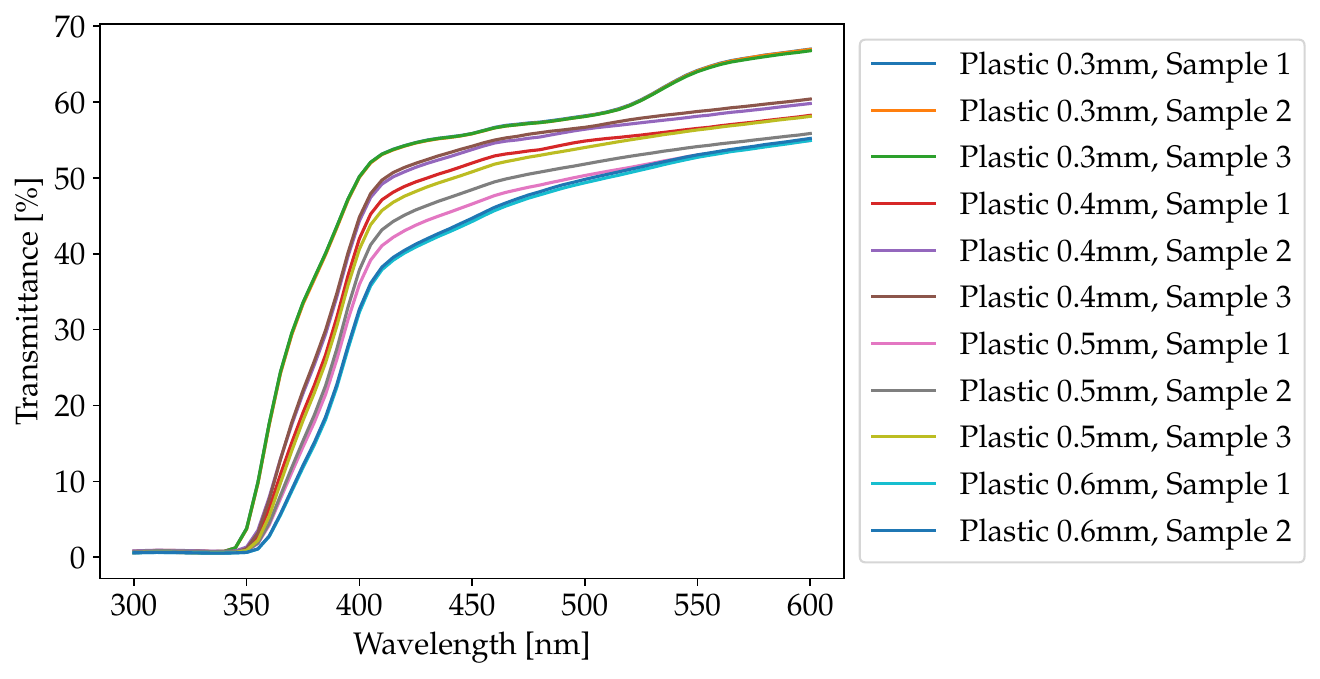}
  \caption{Spectral transmittance measurement of the plastic grid material for different thicknesses \citefig{NDA_thesis}.}
  \label{fig:plastic_grid_trans}
\end{figure}

The measured transmittance spectra for different samples and thicknesses are plotted in Figure \ref{fig:plastic_grid_trans}. As could be expected, the transmittance is getting lower with the increasing thickness of the sample. An average transmittance between 40 and 60\% is measured in the 400-500~nm range. The transmittance in this interval is fitted as a function of the sample thickness to extrapolate the transmittance at the desired thickness\footnote{The thickness of the grid between scintillators is 0.2~mm for most of the channels, and 0.5~mm for the central middle cross separating the 4 quarters of the module \cite{NDA_thesis}.}. The grid having very thin segments, the transmittance over the 3~mm of grid height between two scintillators can be as high as 60\%. Furthermore, the grid is placed at the extremity of the scintillators on the SiPM side. The grid transparency will therefore have a significant contribution to the optical crosstalk and has to be accounted for in the optical simulations.

\begin{figure}[H]
\centering
  \includegraphics[width=0.7\textwidth]{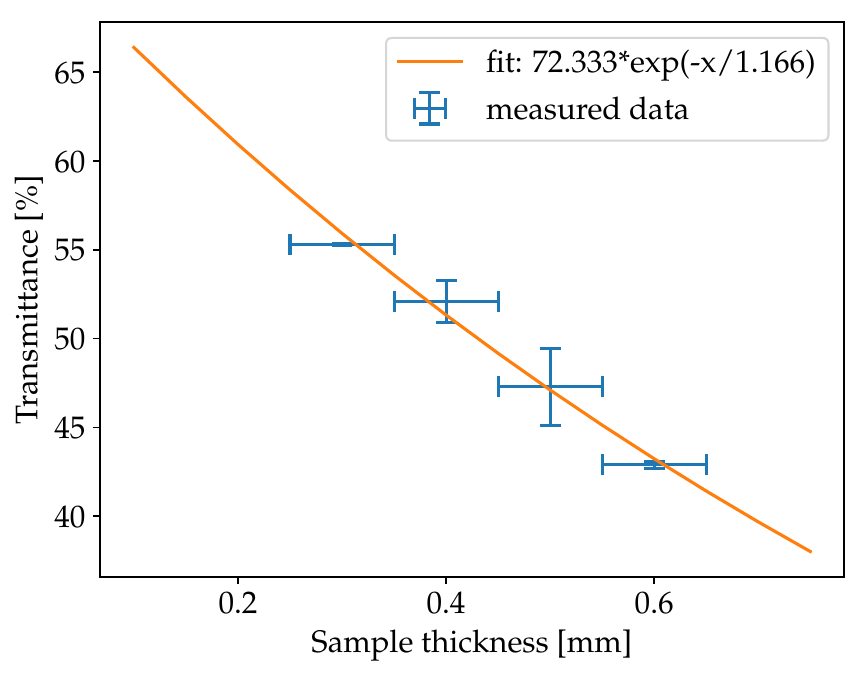}
  \caption{Transmittance of the epoxy resin used to print the plastic alignment grid as a function of the sample thickness, fitted with an exponential function \citefig{NDA_thesis}. An attenuation length of 1.166~mm is found.}
  \label{fig:plastic_grid_trans_fit_vs_thckness}
\end{figure}

Because of the complex shape and mechanical fragility of the plastic grids, no reflective or absorbing coating can be uniformly applied to the pieces to reduce the crosstalk from the grid. A last point that should be mentioned is the effect of "yellowing" of the resin that has been observed on the printed plastic grids after some times. This yellowing effect, caused by thermo-oxidation, appears to be quite common for epoxy based resins \cite{epoxy_yellowing}, and has not shown any impact on the mechanical properties of the alignment grid. It is therefore not leading to any issue for the polarimeter module since it does not have a significant effect on the transmittance properties of the material either.

\subsection{Scintillator bars and the importance of the surface roughness}\label{subsec:scintillator_jacobian_roughness-meas}

As previously mentioned, the scintillator bars' dimensions for POLAR-2 are 5.9$\times$5.9$\times$125~mm$^3$. We describe here the features of the scintillators important for the optical simulation of the module. Two types of scintillators were considered as candidate for POLAR-2: the EJ-248M plastic scintillator from Eljen Technology \cite{Eljen_ej248m}, already used in POLAR, and the EJ-200 scintillator \cite{Eljen_ej200} from the same company, which is reported to have higher scintillation efficiency.\\

\begin{figure}[H]
\centering
  \includegraphics[width=0.7\textwidth]{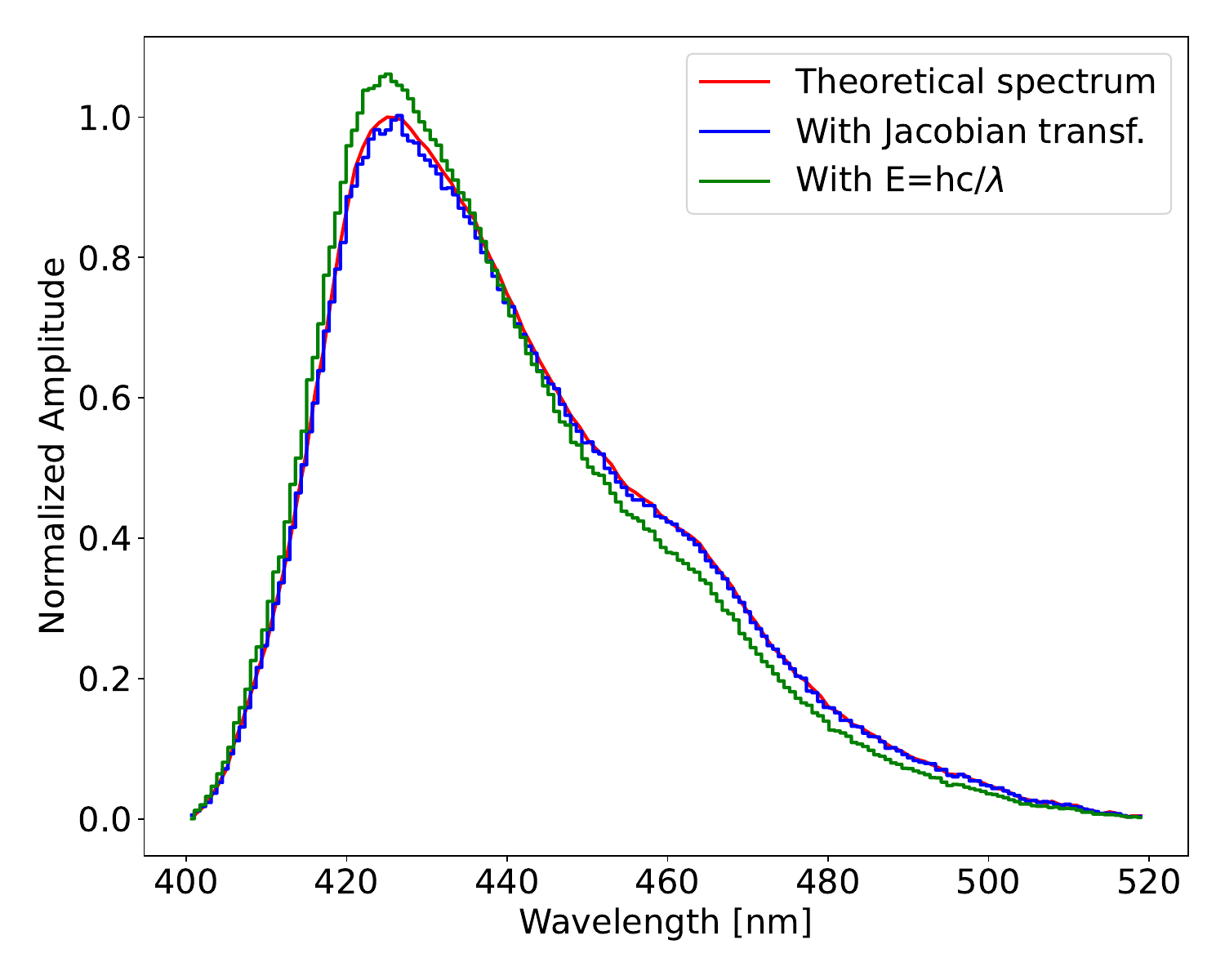}
  \caption{Scintillation spectrum obtained from the simulations compared to the theoretical one used as an input (in red). The incorrect way of converting the emission spectrum simply using the formula $E=hc/\lambda$ is shown in green. The blue curve shows the resulting curve form the simulation when applying the proper Jacobian transformation to convert the wavelength spectrum into an energy spectrum. A very good match is observed between the simulated scintillator emission spectrum and the original one when the Jacobian transformation is properly applied.}
  \label{fig:scintillator_emission_spec_sim}
\end{figure}

First, the emission spectrum of the scintillator has to properly be implemented in the Geant4 optical simulations of the polarimeter\footnote{The optical simulations of a POLAR-2 module will be describe in details in Section \ref{sec:opt_sim}}. The theoretical optical emission spectrum provided in the data sheet is used as an input since it is consistent with the one measured in the lab \cite{POLAR-2_scint_paper}. It should be noted that while the emission of the scintillator is given as a function of the photon wavelength in the scintillator data sheets \cite{Eljen_ej200, Eljen_ej248m}, it has to be implemented versus energy in the optical simulations. As the energy is inversely proportional to the wavelength of a photon ($E=hc/\lambda$), a Jacobian transformation \cite{Jacobian-trans-spectrum} has to be applied for correctly implementing the emission spectrum. The x-axis of the spectrum has therefore to be transformed as follows:

\begin{equation}
    \dd\lambda \rightarrow \dd E = -\frac{hc}{\lambda^2}\dd\lambda,
\end{equation}

while the y-axis of the spectrum becomes:

\begin{equation}
    f(\lambda)\dd\lambda=f(E)\dd E \implies f(E) = f(\lambda)\dv{\lambda}{E} = -f(\lambda)\frac{\lambda^2}{hc}.
\end{equation}

where $f(\lambda)$ is the spectrum as a function of the wavelength, provided in the data sheet, and f(E) the spectrum versus energy as implemented in the optical simulations. The negative sign just represents the opposite direction of integration between $\lambda$ and E. This transformation also applies to other optical spectra implemented in the simulations, like the PDE of the SiPM. The importance of correctly dealing with input spectra is depicted in Figure \ref{fig:scintillator_emission_spec_sim}.

The EJ-200 scintillator type has been studied as a potential replacement for the previously used EJ-248M in POLAR since it is reported to have a higher scintillation efficiency: 10'000~photons/MeV versus 9'200~photons/MeV for EJ-248M. EJ-200 has therefore been used for the first prototype modules of POLAR-2, but showed a lower efficiency than EJ-248M when testing both modules with radioactive sources. This observation is contrary to what we could expect from the EJ-200 higher scintillation efficiency. Discussions with the manufacturer lead to the idea that the EJ-200 material being softer, its surface quality might be lower when diamond milling the scintillators, degrading the final optical efficiency. In order to confirm the theory that a rougher surface for EJ-200 could imply more photons lost at the scintillator interface, the surface quality of both types of scintillators had to be characterized. For this purpose, we used an Interference Optical Microscope (IOM) developed at the Geneva High School of Landscape, Engineering and Architecture (HEPIA) \cite{JOBIN2008896, JOBIN_SPIE}.\\

%As a first trial, a Keyence VHX optical microscope has been used to measure the surface topology of bar samples (see Figure \ref{fig:Kexence_ej248m}). The scintillators being transparent, using an optical microscope was not precise enough to map the surfaces of both EJ-200 and EJ-248M and compare their quality.

%\begin{figure}[H]
%\centering
%  \includegraphics[width=\textwidth]{Pictures/scintillators/EJ248M_VHX_surface.jpg}
%  \caption{Surface quality measurement of an EJ-248M scintillator bar using a Keyence VHX optical microscope.}
%  \label{fig:Kexence_ej248m}
%\end{figure}

\begin{figure}[H]
\centering
  \hspace*{-1cm}\includegraphics[width=0.6\textwidth]{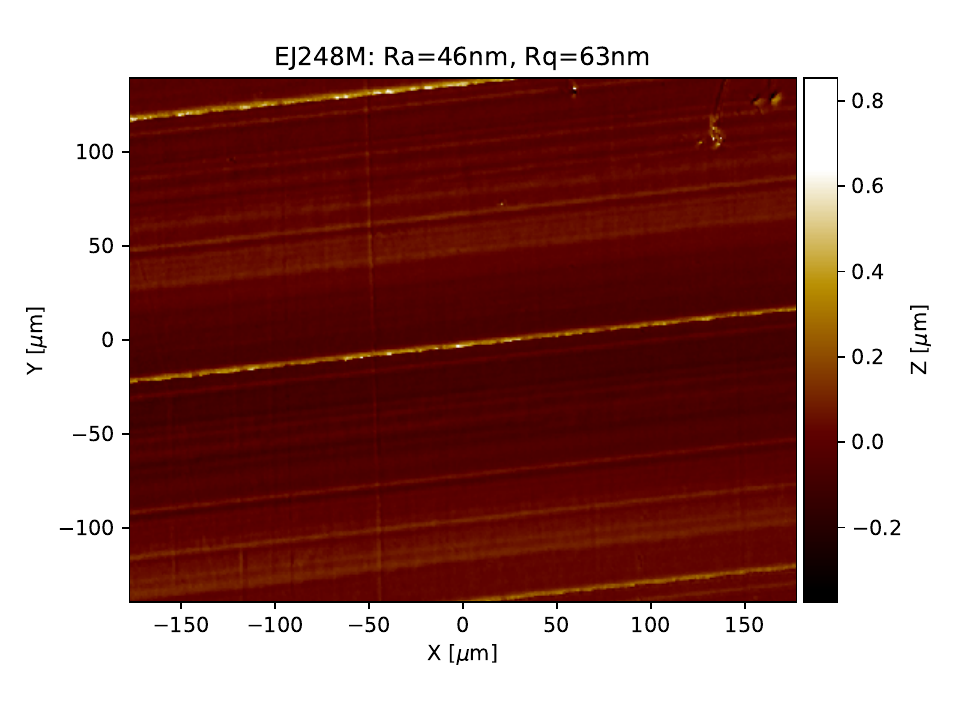}\includegraphics[width=0.6\textwidth]{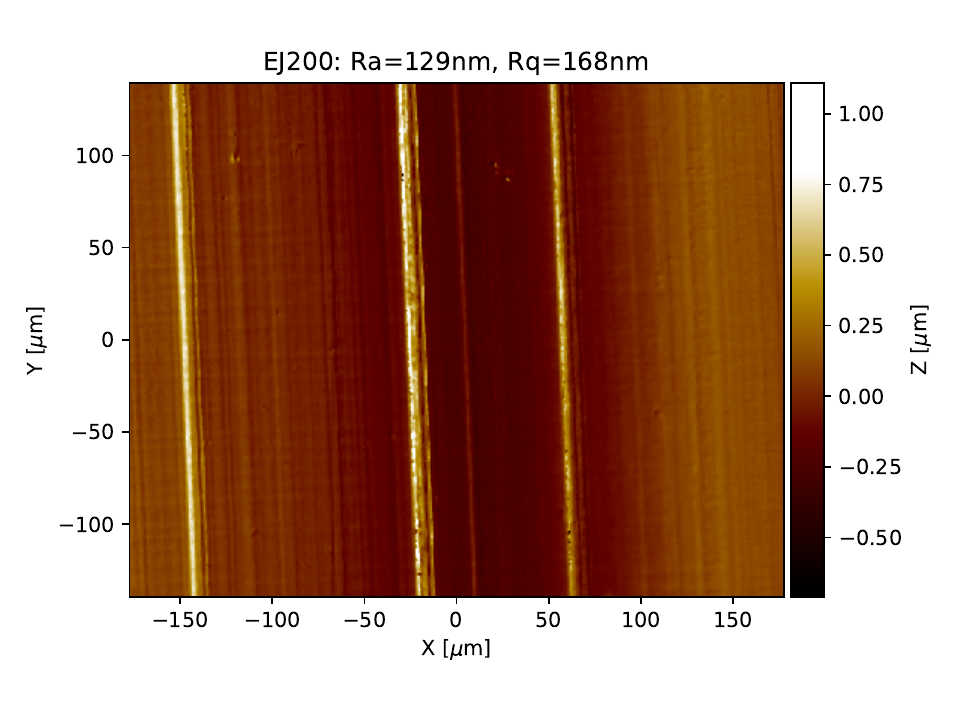}
  \caption{Surface scan measured on a EJ-248M (\textbf{left}) and EJ-200 (\textbf{right}) scintillator bar with the IOM \citefig{NDA_thesis}.}
  \label{fig:ej248m_ej200_IOMmap}
\end{figure}

The IOM setup, described in \cite{JOBIN2008896, NDA_thesis}, is composed of an optical objective read out by a CCD camera, and a support for placing the sample controlled via piezo-electric elements for very fine moving step. The sample surface plane is slightly tilted compared to the microscope focal plane in order to vertically scan the slope of the sample \cite{JOBIN_SPIE}. Examples of a 266$\times$355~$\mu$m area measured with the IOM for both EJ-248M and EJ-200 diamond-polished scintillators are displayed in Figure \ref{fig:ej248m_ej200_IOMmap}. Clear lines due to the diamond polishing process can be observed on the samples. The arithmetic average height $R_a$ and the root mean square roughness $R_q$ \cite{GADELMAWLA2002133} are computed for both scintillators and given at the top of each map. Values of $R_a=46$~nm and $R_q=63$~nm are computed for the EJ-248M scan presented in Figure \ref{fig:ej248m_ej200_IOMmap}, while the EJ-200 scan gives $R_a=129$~nm and $R_q=168$~nm. The EJ-200 scanned surface is clearly rougher than the EJ-248M one. Several samples of both types have been measured similarly in order to gather statistics on the surface quality of both plastics.

The implication of roughness on the optical efficiency of the system will be discussed in section \ref{subsec:roughness_lightoutput}, where the implementation of the measured roughnesses for different scintillator types in the optical simulations will be discussed. Other types of plastic scintillators, not considered for POLAR-2, have also been measured with the IOM setup as it can serve as a useful input for other experiments.

\subsection{Optical coupling pad}\label{subsec:opt_pad_meas}

\begin{figure}[H]
\centering
  \includegraphics[width=0.7\textwidth]{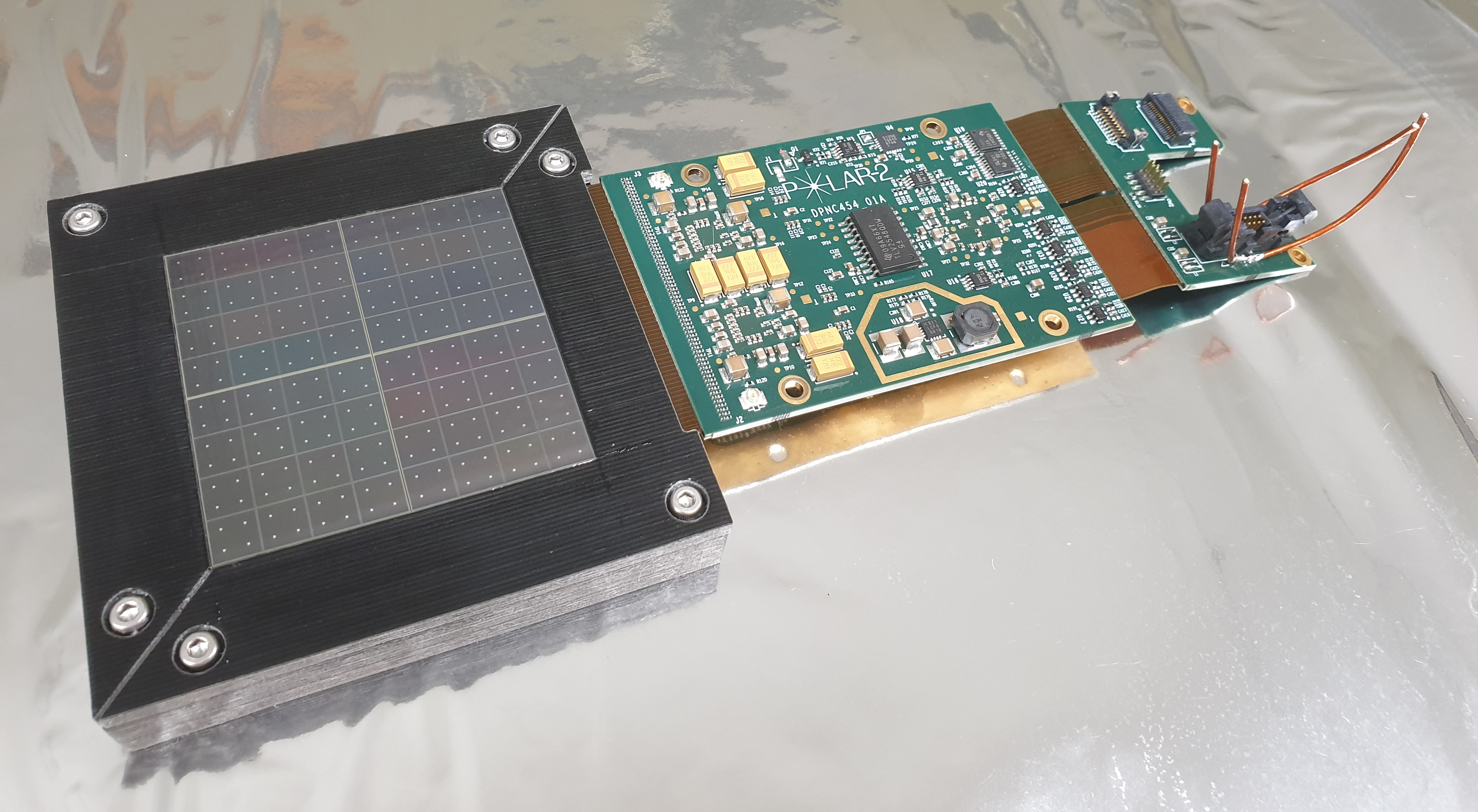}
  \caption{MAPSIL QS1123 optical pad molded on the POLAR-2 FEE \cite{Module_assembly_TN}}
  \label{fig:optical_pad}
\end{figure}

To optically couple the SiPM arrays to the scintillator bars, a silicone-based pad has been developed at DPNC. Made of MAPSIL QS1123 RTV silicone \cite{MAPSIL_paper, MAPSIL_website}, the pad was first developed with thicknesses of 500 and 350~$\mu$m. The technique has later been refined to mold the optical pad directly on the SiPM arrays \cite{Coralie_optpad, Module_assembly_TN}, as displayed in Figure \ref{fig:optical_pad}, and to reduce the pad thickness down to $150^{+0}_{-20}$~$\mu$m.

The importance of having an optical coupling pad to get a smoother contact between the sensors and scintillators compared to directly pressing these two against each other has been shown via optical simulations. A module has been simulated with a 0.3~mm thick optical pad and with an air gap of the same thickness. Injecting 10'000 50~keV electrons in one of the central channels (\#36) of the module, the case with an air gap (872'736$\pm$934 events) gives only 80.06$\pm$0.11\% of the photons detected in the case where an optical pad was used (1'090'087$\pm$1'044 events). This 0.3~mm thick air gap might seem dramatic, but it is actually representative of the fact that not all the bars' extremities are in the same plane. The main role of the optical pad is therefore also to correct for misalignment of the scintillator bars in the Z direction (axis along the length of the scintillator). The pad also protects the SiPM surface against vibration-induced mechanical damage during the launch. The silicon material for the optical pad was chosen to have a refractive index comprised between that of the plastic scintillator and the epoxy resin of the SiPM ($n_{scint}=1.58>n_{optpad}=1.5324>n_{SiPM resin}=1.51$) to optimize the light transmission. An abrupt drop in refractive index from the plastic directly to air/vacuum would increase internal reflections, which we want on the Vikuiti sides but not at the interface with the SiPMs where the light should go out of the scintillator bar.

\begin{figure}[H]
\centering
  \includegraphics[width=0.7\textwidth]{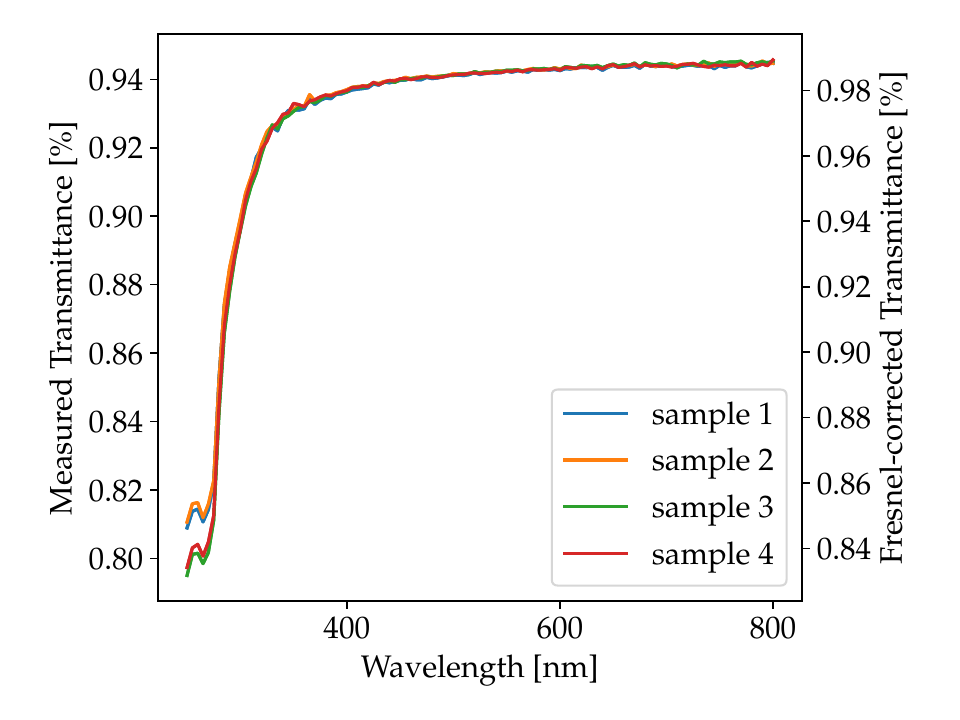}
  \caption{Optical pad spectral transmittance measured with several 350~$\mu$m thick samples \citefig{NDA_thesis}. The right y-axis shows the actual transmittance of the material itself once correcting for Fresnel losses.}
  \label{fig:optical_pad_trans}
\end{figure}

As all the optical photons being collected by the SiPMs will first go through the optical pad, it is imperative to characterize its transmittance. Such measurement was performed with the same setup used for the reflective foils and alignment grid, and is shown in Figure \ref{fig:optical_pad_trans} for several samples of the same thickness. The transmittance measured in the 400-500~nm interval is consistent between samples and its averaged value $93.95\pm0.14$\%. A cutoff is observed below 300~nm, which is a typical feature of Phenyl based silicone \cite{dow_slides_optpad}. The actual optical pad in the final design being 150~$\mu$m thick, the photon loss could be even smaller that what was measured for the 350~$\mu$m thick sample, leading to a higher transmittance. However, most of the photon loss in this transmittance measurement is caused by Fresnel losses at the interface of the sample due to difference in refractive index between the sample and the surrounding air. The measured transmittance can be corrected for Fresnel losses by applying a normalization defined as:

\begin{equation}\label{eq:Fresnel_loss}
    T = 1 - \qty(\frac{n_{air}-n_{optpad}}{n_{air}+n_{optpad}})^2
\end{equation}

where $n_{optpad}=1.5324$ and $n_{air}=1.000293$ are the refractive index of the optical pad and of the surrounding air, respectively. We here use the fact that the optical beam used to measure the transmittance is orthogonal to the sample plane. The corrected transmittance in the 400-500~nm interval is $98.29\pm0.14$\%, which matches the value reported by CNES \cite{MAPSIL_paper} for this material. One can note that the Fresnel losses in the actual polarimeter module are not so large since the refractive index of the optical pad and that of the surrounding materials (scintillator bars and SiPM's resin) are relatively comparable.

\newpage
\section{Module's optical simulations with Geant4 and calibration}\label{sec:opt_sim}

\begin{figure}[H]
\centering
  \includegraphics[width=0.7\textwidth]{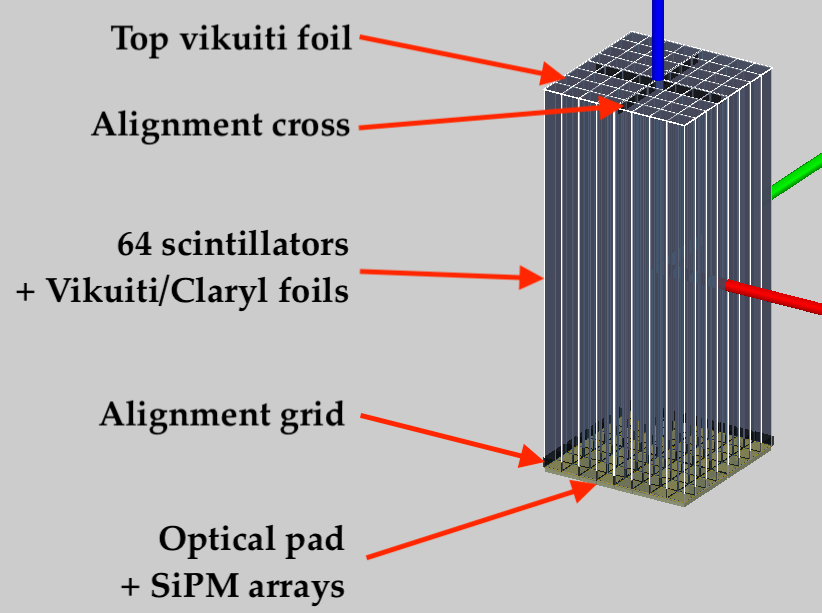}
  \caption{Implementation of the polarimeter module in the optical simulations \citefig{NDA_thesis}.}
  \label{fig:optsim_target}
\end{figure}

The optical simulation of the polarimeter module \cite{Git-code}, whose Geant4 visualization rendering is displayed in Figure \ref{fig:optsim_target}, contains all the key optical components to fully simulate the optical behavior of the system. It is composed of 64 scintillator bars, individually wrapped with 4 strips of Vikuiti surrounded by a foil of Claryl. A 50$\times$50~mm$^2$ Vikuiti foil is placed at the top of the target. The alignment plastic grid is built at the other extremity of the module, where the optical pad and SiPM arrays are also placed. Environmental volumes, whose material can either be assigned to \textit{Vacuum} to reproduce the behavior in space or \textit{Air} to reproduce the behavior of tests performed on ground, are built between each elements which are not physically in contact. This is for instance the case of the Vikuiti strips and scintillator bars, which are separated by such volumes to properly implement the properties of the optical interfaces of each compounds. An air/vacuum gap is needed between the scintillators and the Vikuiti foils in order to have a big difference in refractive index between the bar and its surrounding material to maximize internal reflections. Gluing the reflective foils directly on the scintillators would deteriorate the light yield as it would decrease the amount of internal reflections in the bar. General peculiarities of implementing optical simulations in Geant4 are discussed in \cite{G4Opt_peculiarities}, which was useful for the work presented here.

\subsection{Energy dependent scintillation: the Birks effect}

The scintillation efficiency, that is the amount of light produced per unit of deposited energy, is a crucial parameter to fully understand the optical response of the module. The scintillation efficiency of both EJ-248M and EJ-200 scintillators is given in the data sheet for 10~MeV electrons \cite{Eljen_ej200, Eljen_ej248m}, but its value is energy and particle dependent. Indeed, the efficiency drops at lower energy, as described by the Birks' law \cite{Birks1951}:

\begin{equation}\label{eq:Birks_law}
    \dv{L}{x}=\frac{S \dv{E}{x}}{1+kB \dv{E}{x}}
\end{equation}

where L is the light yield of the scintillator, S is its scintillation efficiency, $\dv{E}{x}$ the energy loss per track length, and kB is the Birks' constant in units of distance over energy. For our scintillator candidates this constant is\footnote{The Birks constant was measured for EJ-248M in \cite{ZHANG201594} for POLAR, the same value is assumed for EJ-200 as the material composition is very similar.} $kB = 0.143$~mm/MeV \cite{ZHANG201594} (with a density of $1.023$~g/cm$^3$) and is given as an input to the optical simulations. This is typical of polyvinyl toluene (PVT) materials, for which the Birks' constant is around $1.5\cdot 10^{-2}$~g MeV$^{-1}$cm$^{-2}$ \cite{Birks1964_book, Knoll_book, TORRISI2000523}. Interesting limits of this law can be noticed:

\begin{equation}
\lim_{\dv{E}{x}\to 0} \dv{L}{x} = S \dv{E}{x} \;;\qquad \lim_{\dv{E}{x}\to \infty} \dv{L}{x} = \frac{S}{kB} = cst.
\end{equation}

The luminescence yield can also be expressed per unit of energy as follows:

\begin{equation}
    \dv{L}{E}=\frac{S}{1+kB \dv{E}{x}}
\end{equation}

\begin{figure}[H]
\centering
  \hspace*{-1cm}\includegraphics[height=0.45\textwidth]{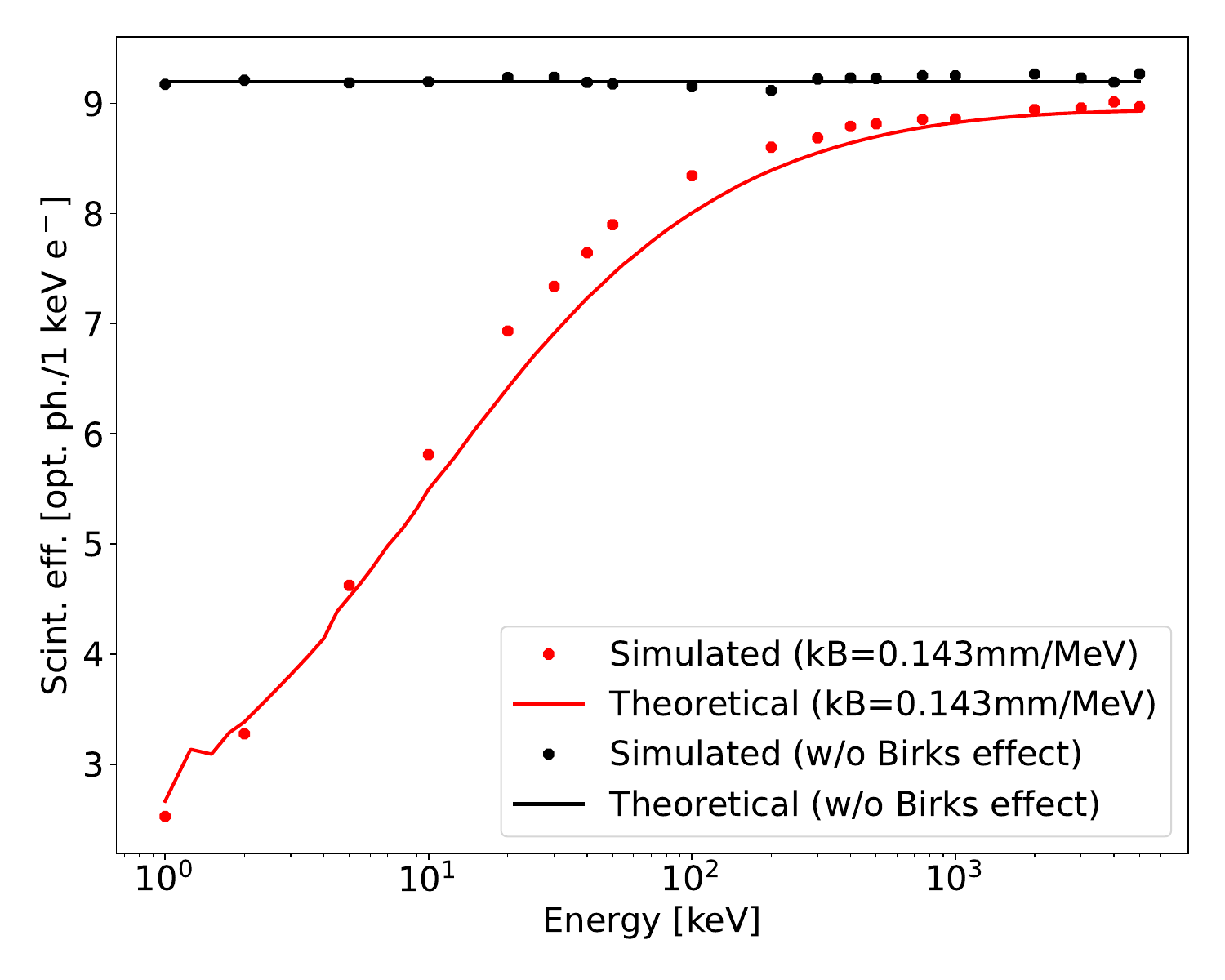}\includegraphics[height=0.45\textwidth]{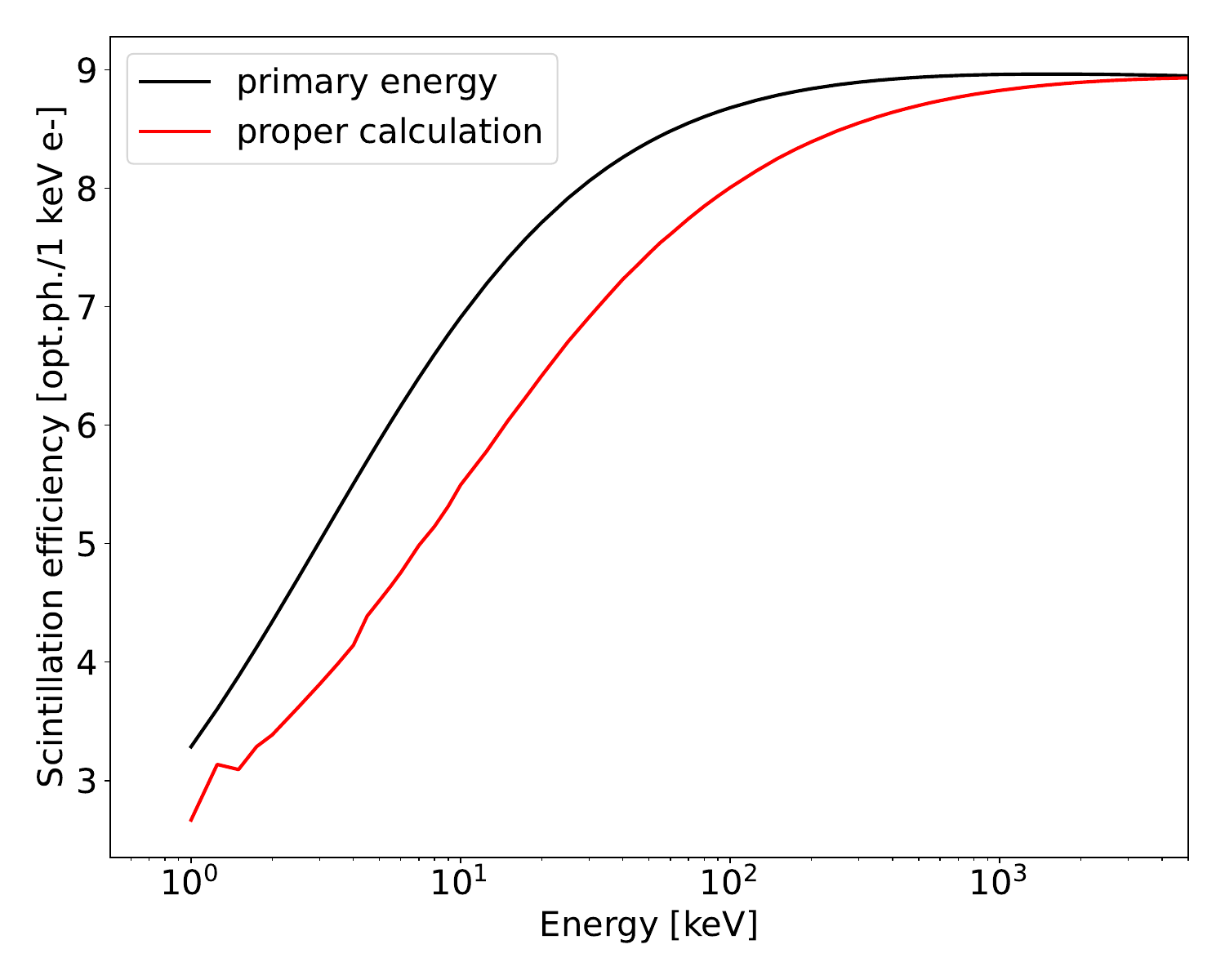}
  \caption{\textbf{Left:} Simulated scintillation efficiency as a function of energy compared to the theoretical behavior based on Birks' law, with and without Birks effect. In the case with no Birks effect (kB is set to zero), no energy dependence of the nominal 9.2 photons/keV scintillation efficiency is observed. \textbf{Right:} Comparison of the theoretical Birks effect scintillation efficiency as a function of energy when accounting only the primary energy of the electron versus properly calculating the electron's energy as it looses energy in the scintillator. \citefigadapt{NDA_thesis}}
  \label{fig:birks_simulation}
\end{figure}

The scintillation efficiency in optical photons per keV is plotted in red in Figure \ref{fig:birks_simulation} as a function of energy for incoming electrons both from the theoretical Birks formula \eqref{eq:Birks_law} and from the optical simulation results. The simulations slightly over-estimates the scintillation efficiency in the 10-300~keV range compared to the theoretical value. The simulated curve not accounting for Birks' effect is also shown in black and matches with the constant 9.2 optical photons per keV provided in the data sheet. Figure \ref{fig:birks_simulation} also shows the comparison of two methods used to compute the theoretical behavior of Birks' effect on the scintillation efficiency.

\begin{figure}[H]
\centering
  \includegraphics[width=0.7\textwidth]{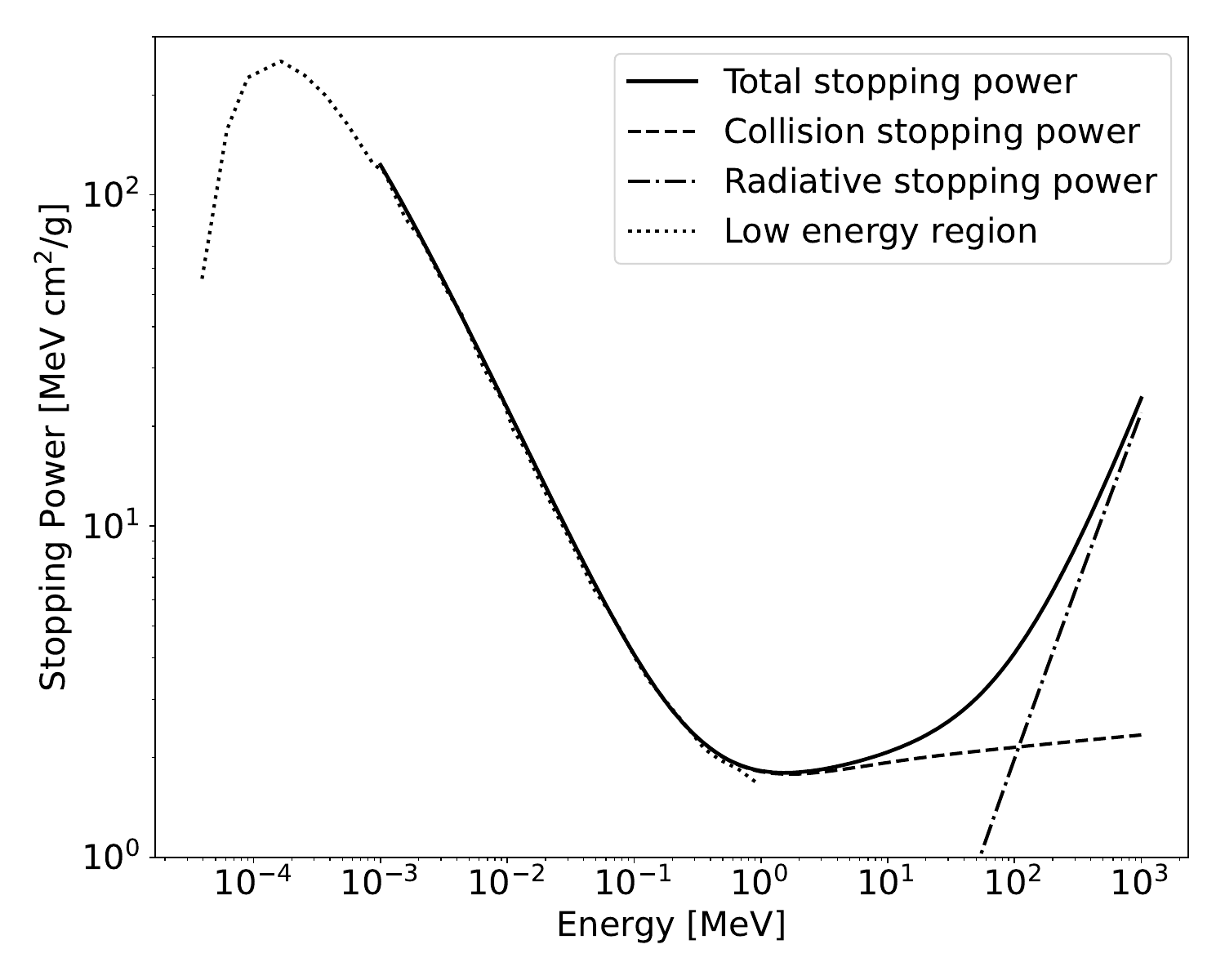}
  \caption{Electron stopping power in polyvinyl toluene \citefigadapt{NDA_thesis}. The sub-keV part is taken from \cite{FRANCIS20112307}, and the high energy part from \cite{ESTAR-database}.}
  \label{fig:electron_stopping_power}
\end{figure}

The simplest but incorrect way of making the computation is to only account for the primary energy of the electron, read the stopping power $\dv{E}{x}$ in PVT in Figure \ref{fig:electron_stopping_power} for the given energy, and use the expression \eqref{eq:Birks_law} to simply compute the scintillation efficiency for each energy. The correct way of computing the scintillation efficiency for each energy is to start from the primary energy, read the electron energy loss in Figure \ref{fig:electron_stopping_power}, apply this energy loss on an infinitesimal track length to compute a new energy, and repeat the process until the electron has deposited all its energy. Looping down over the energy of the electron as it deposits part of it in the scintillator is the proper way of calculating the scintillation efficiency, and is compared to the over-estimated scintillation efficiency got from using the primary energy only in the right plot of Figure \ref{fig:birks_simulation}.

\subsection{Influence of the scintillator surface quality on the light output}\label{subsec:roughness_lightoutput}

After having measured the surface quality of both EJ-200 and EJ-248M scintillators (see Section \ref{subsec:scintillator_jacobian_roughness-meas}), it has to be implemented in the Geant4 optical simulations. The way Geant4 deals with the roughness of a given transparent surface is by deviating each incoming photon when it crosses the surface by an angle $\alpha$ added on top of the refracting angle. This angle is picked from a Gaussian distribution. The input given to the simulations is the width of this Gaussian distribution, $\sigma_\alpha$. The wider the distribution, the rougher the surface. A rougher surface will therefore cause a bigger spread of the photons' deviation. The surfaces can also be seen as composed of many micro-facets \cite{NILSSON201515}, whose orientations are tilted by an angle $\alpha$ compared to the average surface.

\begin{figure}[H]
\centering
  \includegraphics[width=0.6\textwidth]{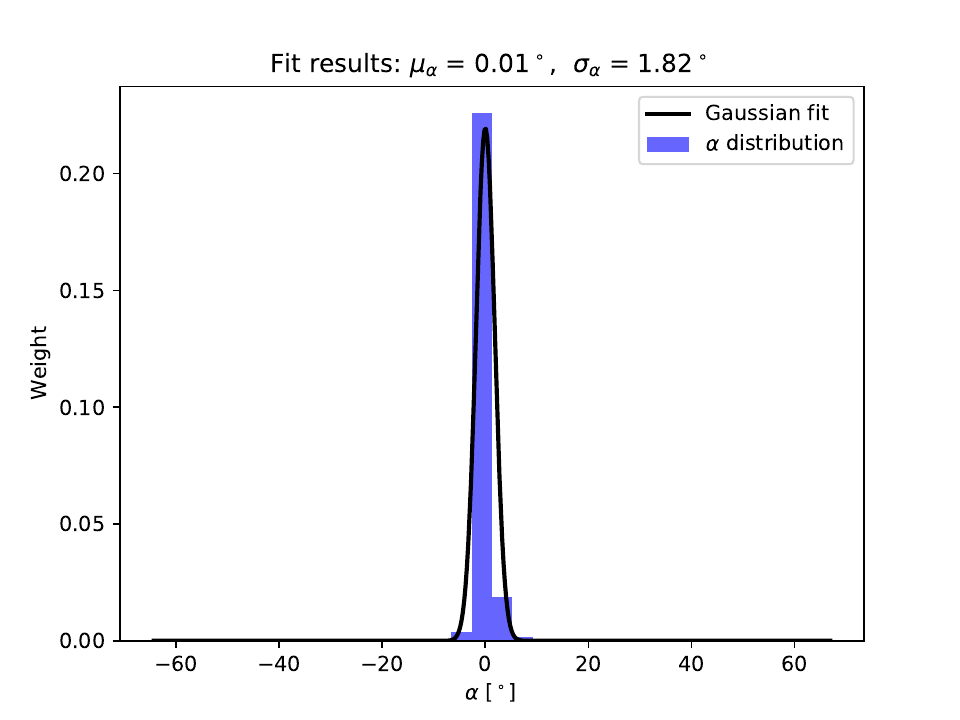}\includegraphics[width=0.6\textwidth]{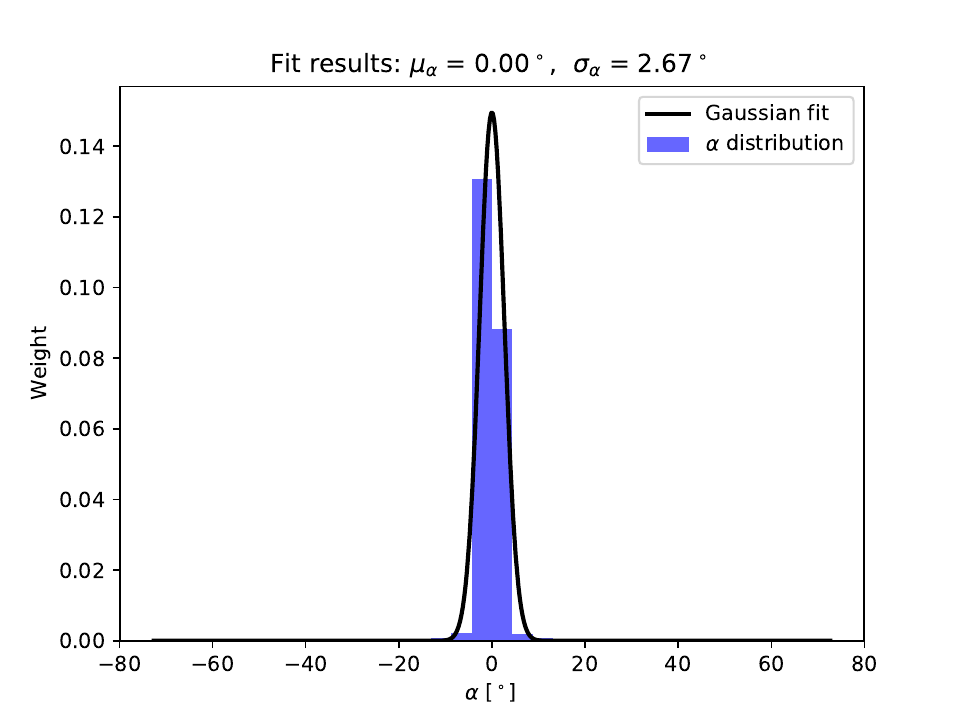}
  \caption{Extraction of the $\sigma_\alpha$ parameter for the samples measured in Figure \ref{fig:ej248m_ej200_IOMmap} \citefig{NDA_thesis}. \textbf{Left:} EJ-248M \textbf{Right:} EJ-200}
  \label{fig:ej248m_ej200_sigalpha}
\end{figure}

The maps measured were analyzed on order to extract a $\sigma_\alpha$ value for each type of plastic (see Section \ref{subsec:scintillator_jacobian_roughness-meas}). A distribution of the angle between two points of the measured map and the average surface plane is filled for every point on the map. Examples of obtained Gaussian distributions for both types of plastics are shown in Figure \ref{fig:ej248m_ej200_sigalpha}. Several maps were measured for each plastic giving the averaged values of $\sigma_\alpha^{EJ-248M} = 1.82\pm0.09^\circ$ and $\sigma_\alpha^{EJ-200} = 3.45\pm0.14 ^\circ$, the latter being bigger since EJ-200 is rougher than EJ-248M. Measured roughness parameters of other types of plastic scintillators are provided in Appendix \ref{sec:appendix_roughness}, as they can be useful for other experiments.\\

We now define the light output fraction as being the fraction of the injected optical photons reaching the SiPMs. This fraction does not include the PDE of the SiPM and is used as a figure of merit of the optical efficiency of the target. Two attenuation lengths can be defined in order to disentangle the intrinsic scintillator material attenuation from the photons lost because of the surface roughness of the scintillator. We therefore define the Bulk Attenuation Length (BAL) and Technical Attenuation Length (TAL) \cite{Kaplon_TAL_BAL_IEEE, Kaplon_TAL_BAL_paper} as follows:

\begin{equation}\label{eq:BAL_TAL}
    BAL:\; I(x)=A_1 e^{-\frac{x}{\lambda_1}} \;;\qquad TAL:\; I(x)=A_1 e^{-\frac{x}{\lambda_1}} + A_2 e^{-\frac{x}{\lambda_2}}
\end{equation}

The BAL represent the intrinsic material property that is given in the manufacturer data sheets \cite{Eljen_ej200, Eljen_ej248m} as being 380~cm for EJ-200 and 250~cm for EJ-248M. The light output fraction was simulated for the BAL injecting optical photons along the bar direction at different height in the scintillator and for different roughnesses. The corresponding Gun Particle Source (GPS) file used in Geant4 to inject optical photons is:\\[0.2em]

\begin{tabular}{lcl}
{\color{black}\begin{lstlisting}
/gps/particle opticalphoton
/gps/pos/type Plane
/gps/pos/shape Rectangle
/gps/pos/centre 0.0 0.0 1.0 cm
/gps/pos/halfx 0.1 cm
/gps/pos/halfy 0.1 cm
/gps/direction 0 0 -1
\end{lstlisting}}
& \quad &
{\color{black}\begin{lstlisting}
/gps/ene/type Arb
/gps/ene/diffspec 0
/gps/hist/type arb
/gps/hist/point 2.5e-06 0.06295
...
/gps/hist/point 3.09e-06 0.00883
/gps/hist/inter Lin
\end{lstlisting}}\\
\end{tabular}

\begin{figure}[H]
\centering
  \hspace*{-1cm}\includegraphics[width=0.6\textwidth]{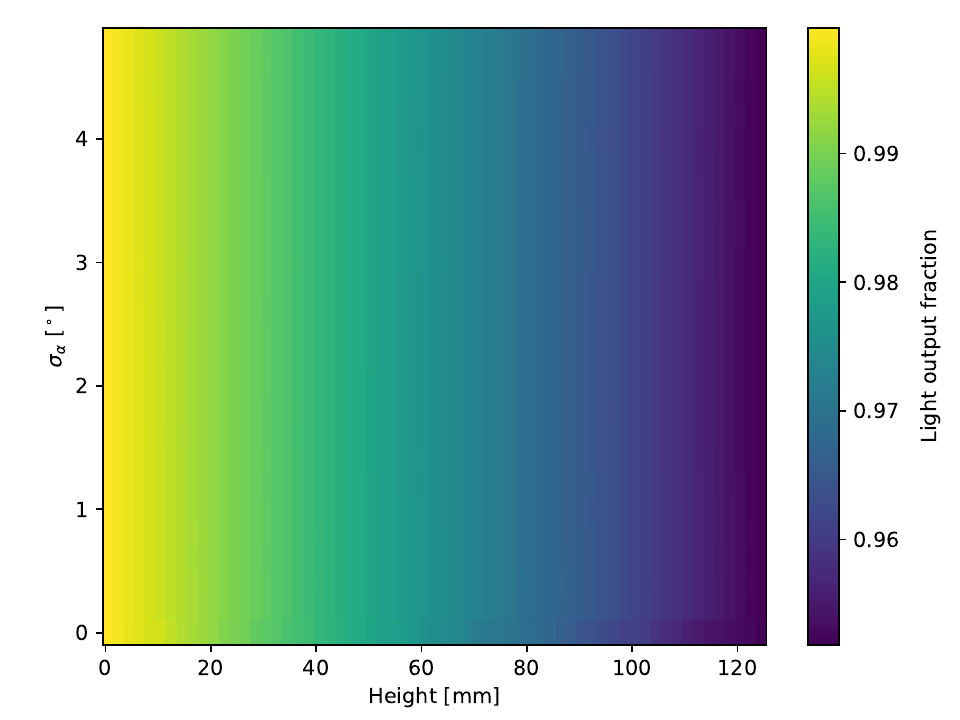}\includegraphics[width=0.6\textwidth]{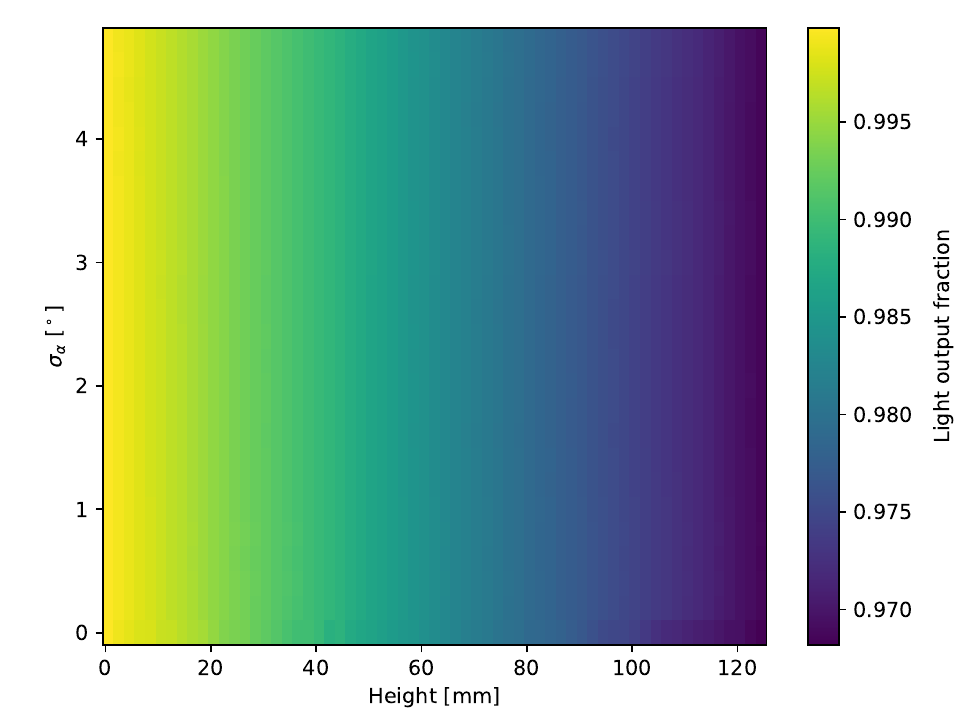}
  \caption{Simulated light output fraction for the EJ-248M (\textbf{left}) and EJ-200 (\textbf{right}) scintillators injecting light parallel to the scintillator length direction \citefig{NDA_thesis}. The simulations are performed with perfect reflectors and 100\% sensitive SiPMs so that the light output only reflects the optical light loss at the scintillator surface due to the roughness. It is shown as a function of the roughness of the scintillator surface and the height at which the light is injected along the scintillator bar, 0~mm being the extremity of the bar on the SiPM side, and 125~mm the extremity of the bar facing the deep space. As the optical light is injected along the scintillator length, the photons are not crossing the 4 long faces of the scintillator bar and are directly going to the sensor. No roughness dependence is therefore observed while the light output decreases when the light is injected further away from the SiPM channel, due to the bulk attenuation length of the scintillating material.}
  \label{fig:ej248m_ej200_lightoutput_straight}
\end{figure}

The resulting light output map from these simulations is shown in Figure \ref{fig:ej248m_ej200_lightoutput_straight}, where we clearly see a decreasing light output with the height for both plastics. This is explained by the bulk attenuation length of the material: when the photons are injected near the SiPMs (small height), more will reach the sensor than when they are injected on the other side of the bar. No significant dependence of the light output on $\sigma_\alpha$ is observed\footnote{We would expect a small dependency in $\sigma_\alpha$ as some photons can bounce back and travel more than once along the bar, but this is not observed here.}, since the photon are propagating along the bar direction and are therefore not crossing the scintillators surfaces except once when they reach the optical pad and sensor. The same simulations can be ran injecting photons from a given height but in isotropic directions from a sphere. This is done by using the following GPS:

{\color{black}
\begin{lstlisting}
/gps/particle opticalphoton
/gps/energy 3 eV
/gps/pos/type Surface
/gps/pos/shape Sphere
/gps/pos/radius 1 mm
/gps/pos/centre 0.325 0.325 0.0 cm
/gps/ang/type cos
\end{lstlisting}}

\begin{figure}[H]
\centering
  \hspace*{-1cm}\includegraphics[width=0.6\textwidth]{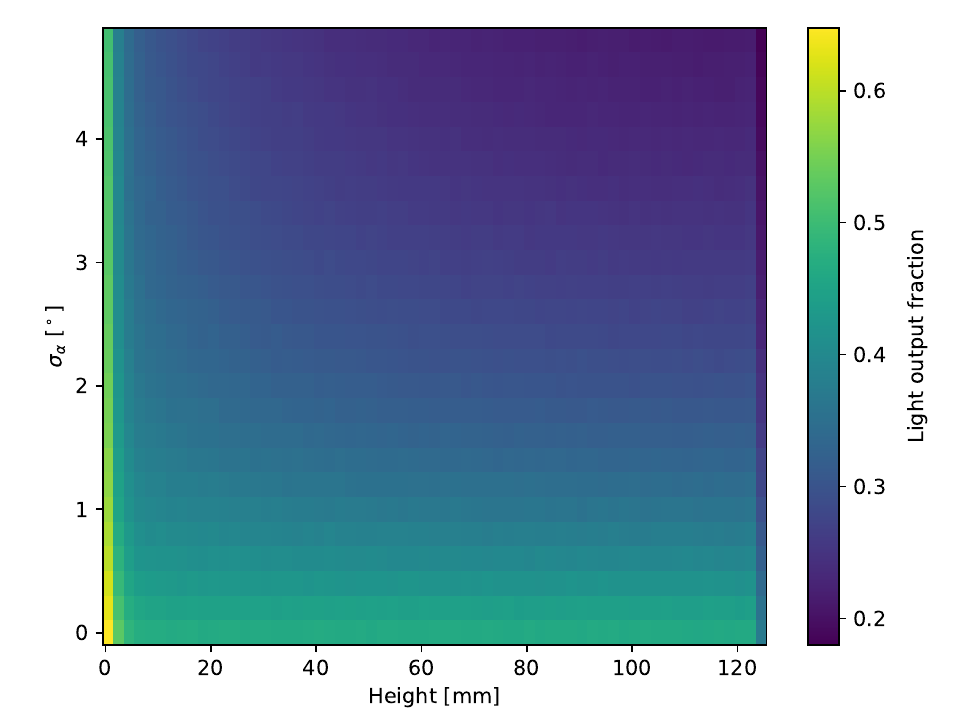}\includegraphics[width=0.6\textwidth]{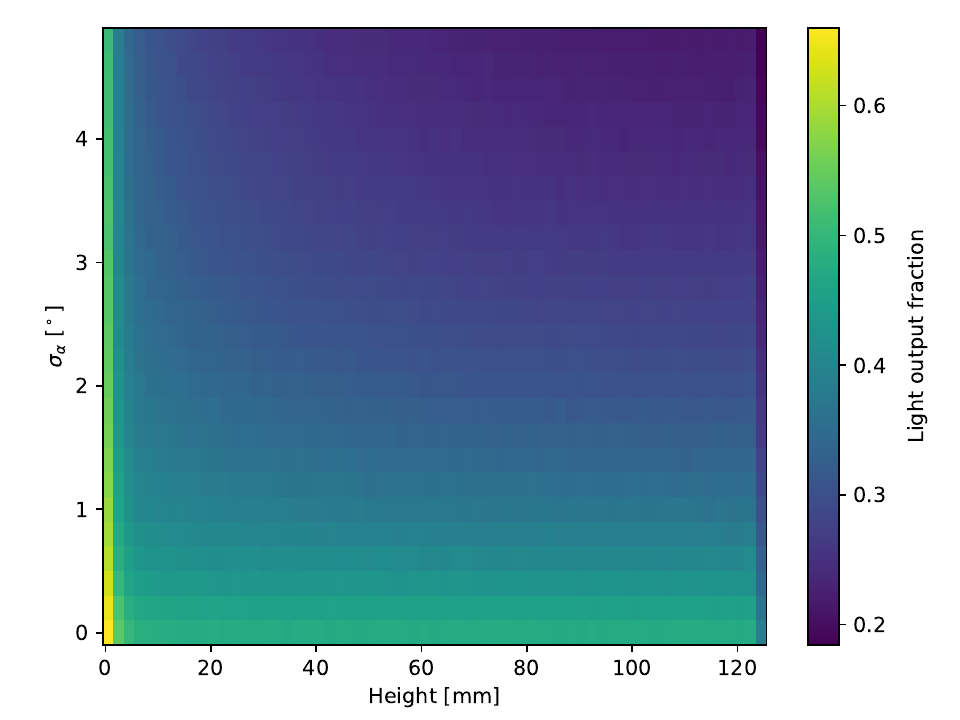}
  \caption{Simulated light output fraction for the EJ-248M (\textbf{left}) and EJ-200 (\textbf{right}) scintillators injecting light with an isotropic symmetry \citefig{NDA_thesis}. The simulations are performed with perfect reflectors and 100\% sensitive SiPMs so that the light output only reflects the optical light loss at the scintillator surface due to the roughness. It is shown as a function of the roughness of the scintillator surface and the height at which the light is injected along the scintillator bar, 0~mm being the extremity of the bar on the SiPM side, and 125~mm the extremity of the bar facing the deep space. As the optical light is injected with a radial symmetry, photons are crossing the scintillator surface and reflected back many times before reaching the sensors. The light output is in consequence highly dependent on the surface roughness, as the loss of photons decreases with a higher surface quality. The light output fraction is also dependent on the injection position, this time not only because of the optical attenuation intrinsic to the material, but also because photons injected further away from the sensors will have crossed the surfaces many more times before reaching it. This reflects the technical attenuation length describe by the second expression in equation \eqref{eq:BAL_TAL}.}
  \label{fig:ej248m_ej200_lightoutput_sphere}
\end{figure}

The resulting light output fraction maps for EJ-248M and EJ-200 are shown in Figure \ref{fig:ej248m_ej200_lightoutput_sphere}. As in the previous case, the light output is decreasing with height. However, this decrease is much more dramatic and is also $\sigma_\alpha$ dependent. As one could expect, the rougher the surface, the lower the light output since more photons are lost when crossing the scintillator surface and being reflected back by the ESR. This reflects the technical attenuation length, described in equation \eqref{eq:BAL_TAL}, which accounts not only for the intrinsic light attenuation from the material but also for the scintillator shape and surface quality, as well as for the performances of the reflectors. This case is therefore more representative of the actual module.

\begin{figure}[H]
\centering
  \includegraphics[width=0.7\textwidth]{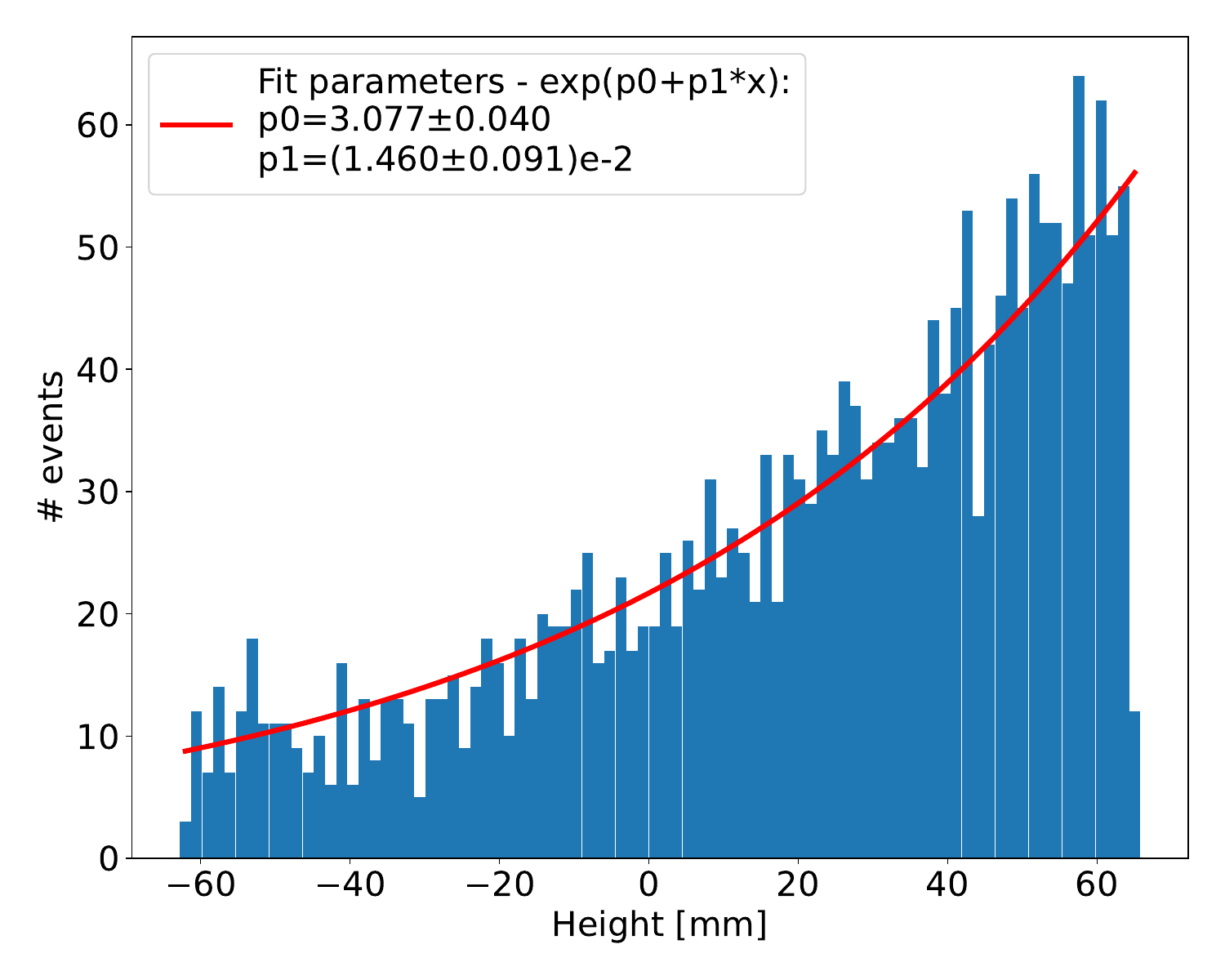}
  \caption{Penetration profile of 60~keV photons in the plastic scintillator fitted with an exponential. The 60~keV energy corresponds to beam energy used to measure the module light yield (see section \ref{sec:LY}).}
  \label{fig:cs137_penetration_profile}
\end{figure}

The light yield of the polarimeter module is measured using 60~keV synchrotron beam (see later in this section). The light output maps in Figure \ref{fig:ej248m_ej200_lightoutput_sphere} can be therefore convolved along the height dimension with the simulated penetration profile of 60~keV photons in the scintillators plotted in Figure \ref{fig:cs137_penetration_profile}.

Through this convolution, we obtain the light output fraction profile as a function of the surface roughness for both scintillator types. The light output fractions can then simply be read from these curves by using the roughness parameters computed from the measurements, $\sigma_\alpha^{EJ-248M} = 1.82\pm0.09^\circ$ and $\sigma_\alpha^{EJ-200} = 3.45\pm0.14 ^\circ$. The corresponding light output fractions for the EJ-248M and EJ-200 scintillators are respectively 0.336$\pm$0.004 and 0.264$\pm$0.004. The impact of the rougher EJ-200 surface is therefore visible on the light output fraction and is countering the higher scintillation efficiency of EJ-200 compared to EJ-248M.

\begin{figure}[H]
\centering
  \includegraphics[width=0.7\textwidth]{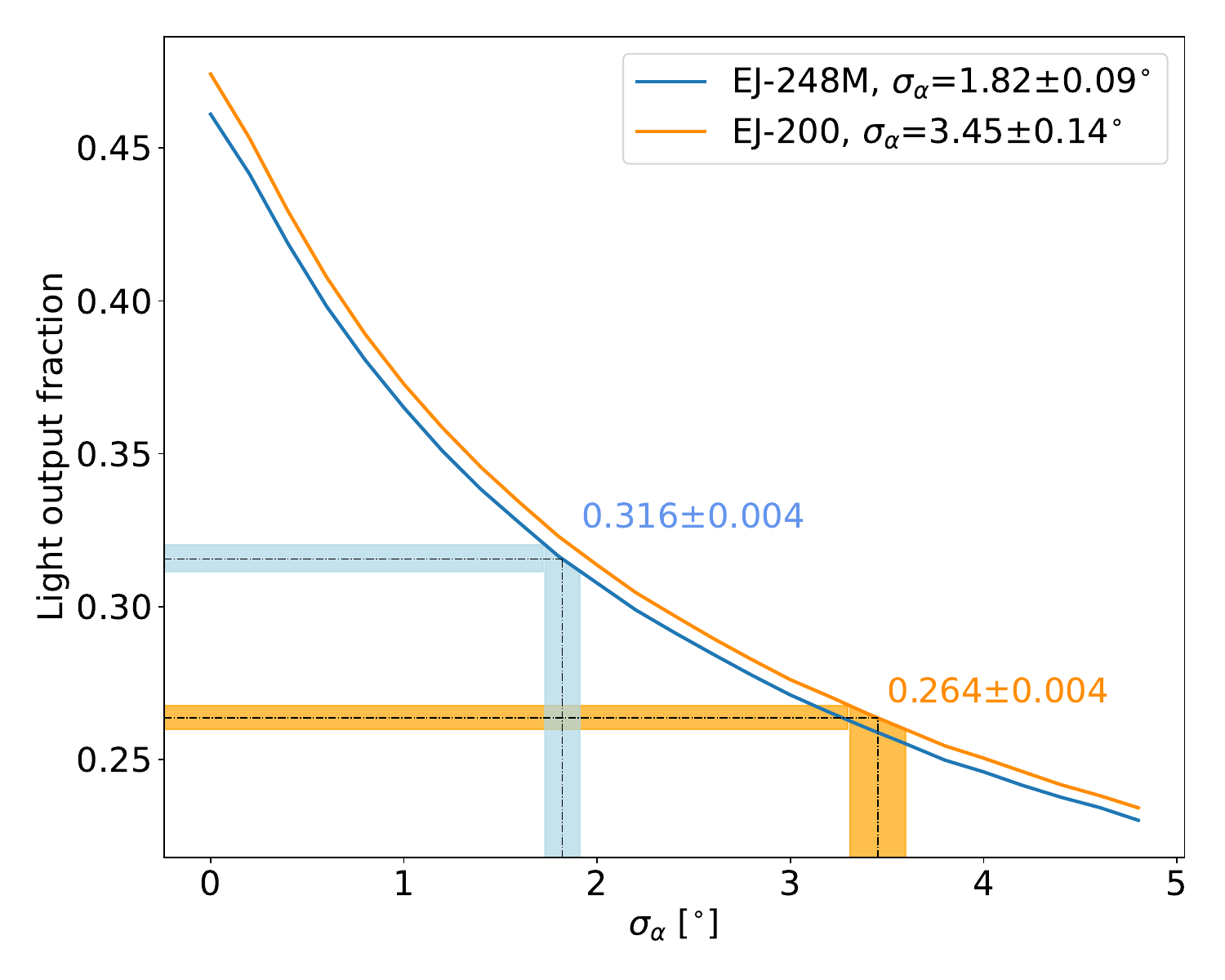}
  \caption{Simulated light output fraction for 60~keV photons for both EJ-248M and EJ-200 plastics as a function of the surface roughness. The light output fractions corresponding to the measured surface roughness are displayed for both scintillator types.}
  \label{fig:lightouput_vs_sig_alpha}
\end{figure}

Furthermore, knowing the $\sigma_\alpha$ values for both types of material from measurements, the height profile of the light output fraction can be extracted from the maps of Figure \ref{fig:ej248m_ej200_lightoutput_sphere}. The resulting curves can be fitted with a sum of two exponentials, as in the TAL expression of equation \eqref{eq:BAL_TAL}.

\begin{figure}[H]
\centering
  \includegraphics[width=0.8\textwidth]{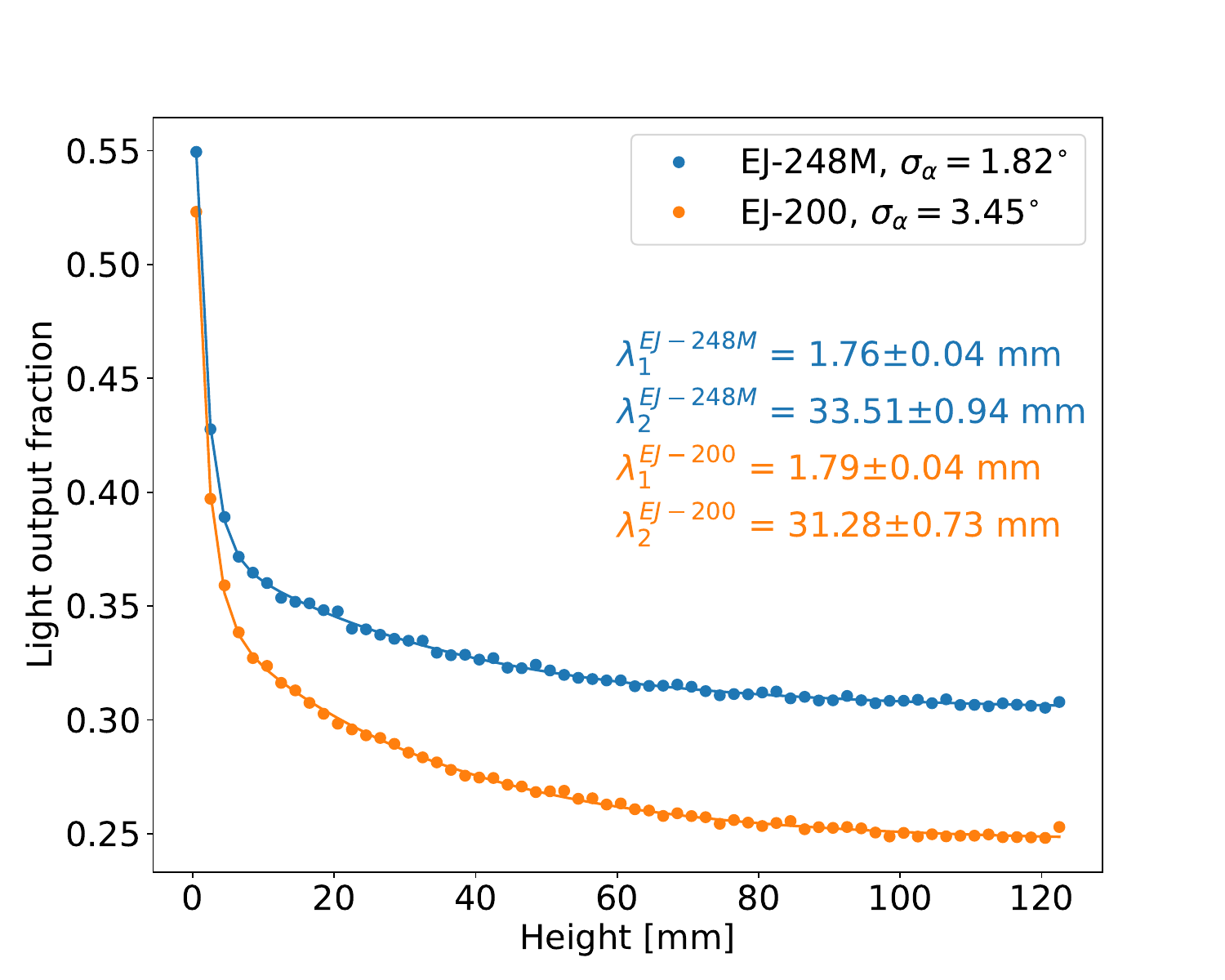}
  \caption{Light output fraction as a function of the injection height for the measured roughness of both EJ-248M and EJ-200 plastics fitted with the TAL expression in equation \eqref{eq:BAL_TAL} \citefig{NDA_thesis}.}
  \label{fig:TAL_exp_fit}
\end{figure}

The effect of scintillator roughness on the light output can further be observed by measuring the light yield of different scintillator shapes. This was done using cylindrically shaped scintillators of 12.7~mm diameter and 25.4~mm height. The light output was in this case larger for EJ-200 than for EJ-248M because of the different volume-to-surface ratio \cite{POLAR-2_scint_paper}. Indeed, with the cylinders having a bigger volume-to-surface ratio, the contribution of the surface roughness to the overall light output was less pro-eminent than for the scintillator bars and was therefore not important enough to counterbalance the bigger scintillation efficiency of EJ-200.

\subsection{Optical light yield}\label{sec:LY}

The light output fraction is a good figure of merit to assess the optical efficiency of the target itself. However, to completely represent the efficiency of the polarimeter module, a more physical quantity is used: the Light Yield (LY). The LY is the number of detected photons, also known as photo-electrons, per unit of incoming energy in the detector. The bigger the LY, the better the sensitivity of the polarimeter, especially at low energies where having a good LY is crucial since only a few photons are produced through the scintillation process.

\begin{figure}[H]
\centering
  \includegraphics[width=0.7\textwidth]{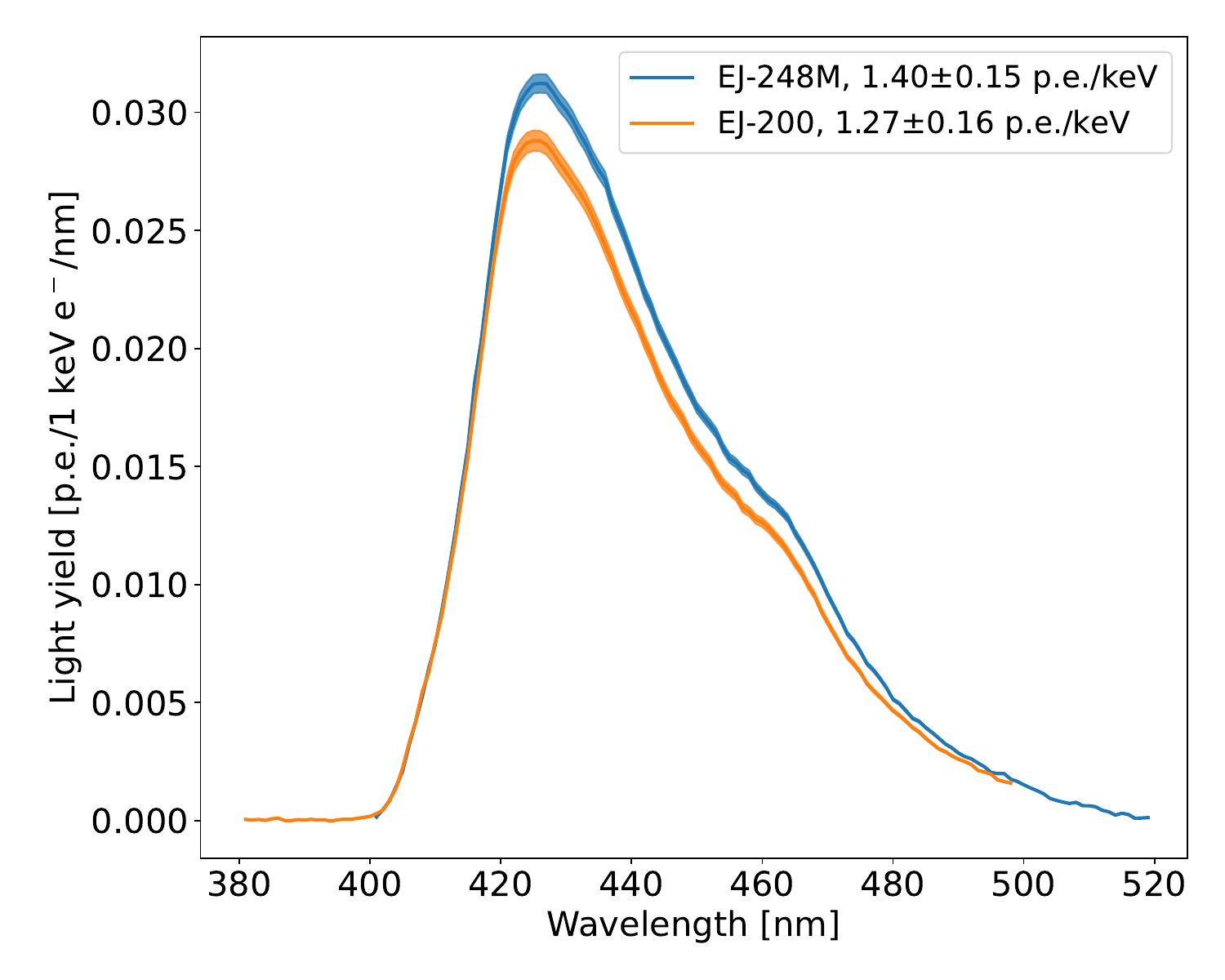}
  \caption{Computed light yield in photoelectrons per keV per unit of wavelength as a function of the photon wavelength for both EJ-248M and EJ-200 scintillators. This light yield spectrum is obtained by convolving the scintillator emission spectrum with the SiPM PDE and the light output fraction for the measured roughness from Figure \ref{fig:lightouput_vs_sig_alpha}. The integration of these curves leads to the module light yields of LY$_{EJ-248M}=1.40\pm0.15$~p.e./keV and LY$_{EJ-200}=1.27\pm0.16$~p.e./keV.}
  \label{fig:spectrum_convlved_ly}
\end{figure}

Based on the light output fractions obtained for both scintillators from simulations, the expected light yield can be computed. This is done in Figure \ref{fig:spectrum_convlved_ly}, where the light output fraction is convolved with the efficiency of the sensors (PDE) and the scintillator emission spectrum. The light yield per unit of wavelength is plotted as a function of the photon wavelength for both plastics. By integrating the curves, the light yield in photo-electrons per keV is obtained. The resulting values are LY$_{EJ-248M}=1.40\pm 0.15$~p.e./keV and LY$_{EJ-200}=1.27\pm 0.16$~p.e./keV. The light yield for EJ-248M is indeed higher than that of EJ-200 although the latter has a larger scintillation efficiency, as observed in the laboratory.

\begin{figure}[H]
\centering
  \includegraphics[width=0.7\textwidth]{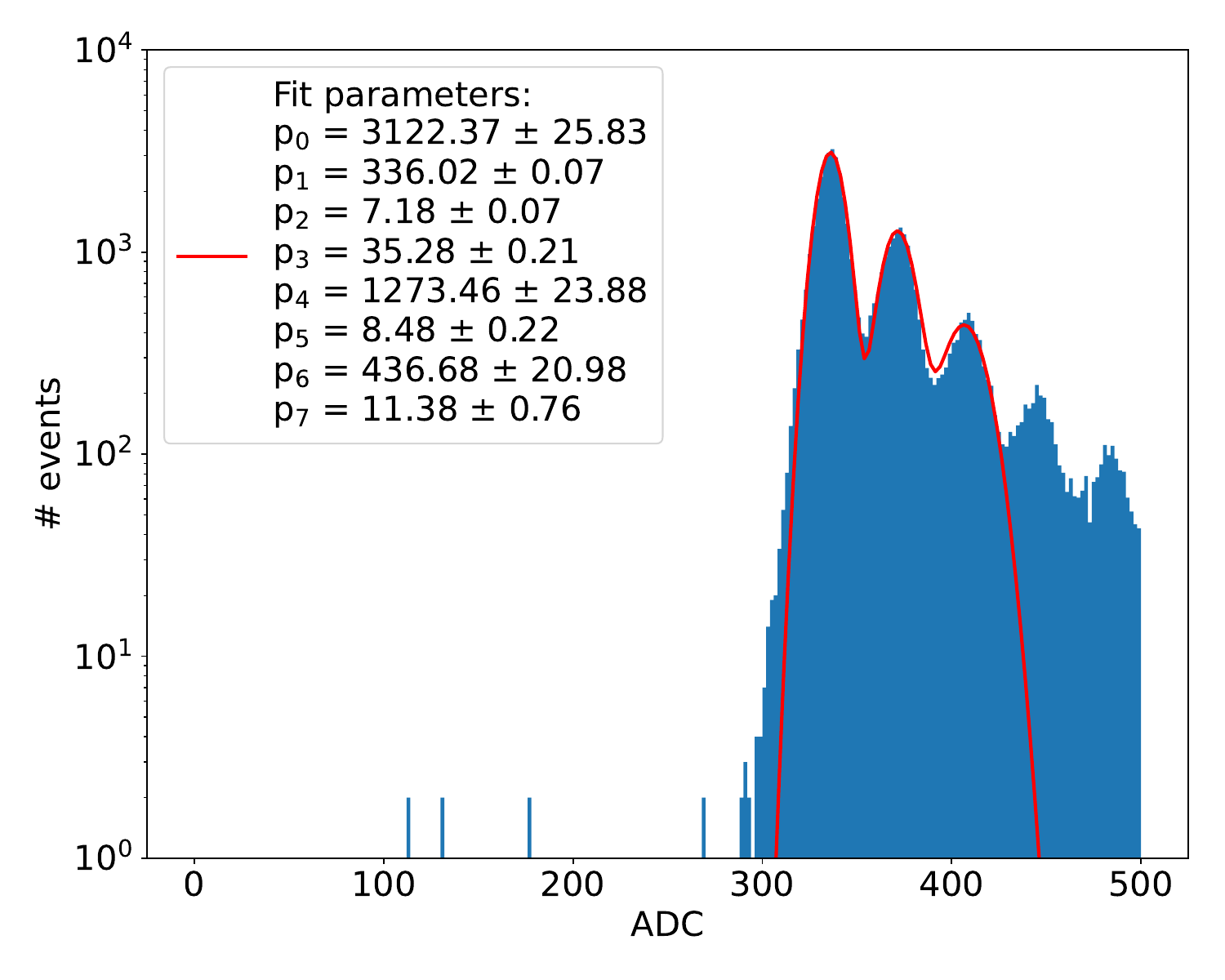}
  \caption{Dark spectrum of the SiPMs measured with a POLAR-2 prototype module \citefig{NDA_thesis}. The first peaks of the finger-like structure are fitted using a sum of Gaussians, as expressed in equation \eqref{eq:finger_fit}. The position of each peak is extracted to determine the distance between peaks, which gives the conversion from HG ADC to photo-electrons.}
  \label{fig:finger_fit}
\end{figure}

The prototype POLAR-2 modules can be calibrated either using radioactive (e.g. $^{241}$Am and $^{137}$Cs) sources in the POLAR-2 laboratory at CERN \cite{NDA_thesis}, or with an X-ray beam like the one at the European Synchrotron Radiation Facility (ESRF) in Grenoble (France), where a calibration campaign with several prototype modules took place in April 2023 \cite{ESRF_data, ESRF_paper}. A dark spectrum is first taken to measure the single photo-electron spectrum, also known as \textit{finger plot}, to determine the conversion between detected photons (or photoelectrons) and ADC counts. The finger structure, of which a measured example is shown in Figure \ref{fig:finger_fit} is fitted using a sum of Gaussian distributions: 

\begin{equation}\label{eq:finger_fit}
\begin{split}
p_0 \exp(-\frac{1}{2}\qty(\frac{x-p_1}{p_2})^2) + p_4 \exp(-\frac{1}{2}\qty(\frac{x-(p_1+p_3)}{p_5})^2)\\ + p_6 \exp(-\frac{1}{2}\qty(\frac{x-(p_1+2 p_3)}{p_7})^2)
\end{split}
\end{equation}

\begin{figure}[H]
\centering
  \includegraphics[width=0.7\textwidth]{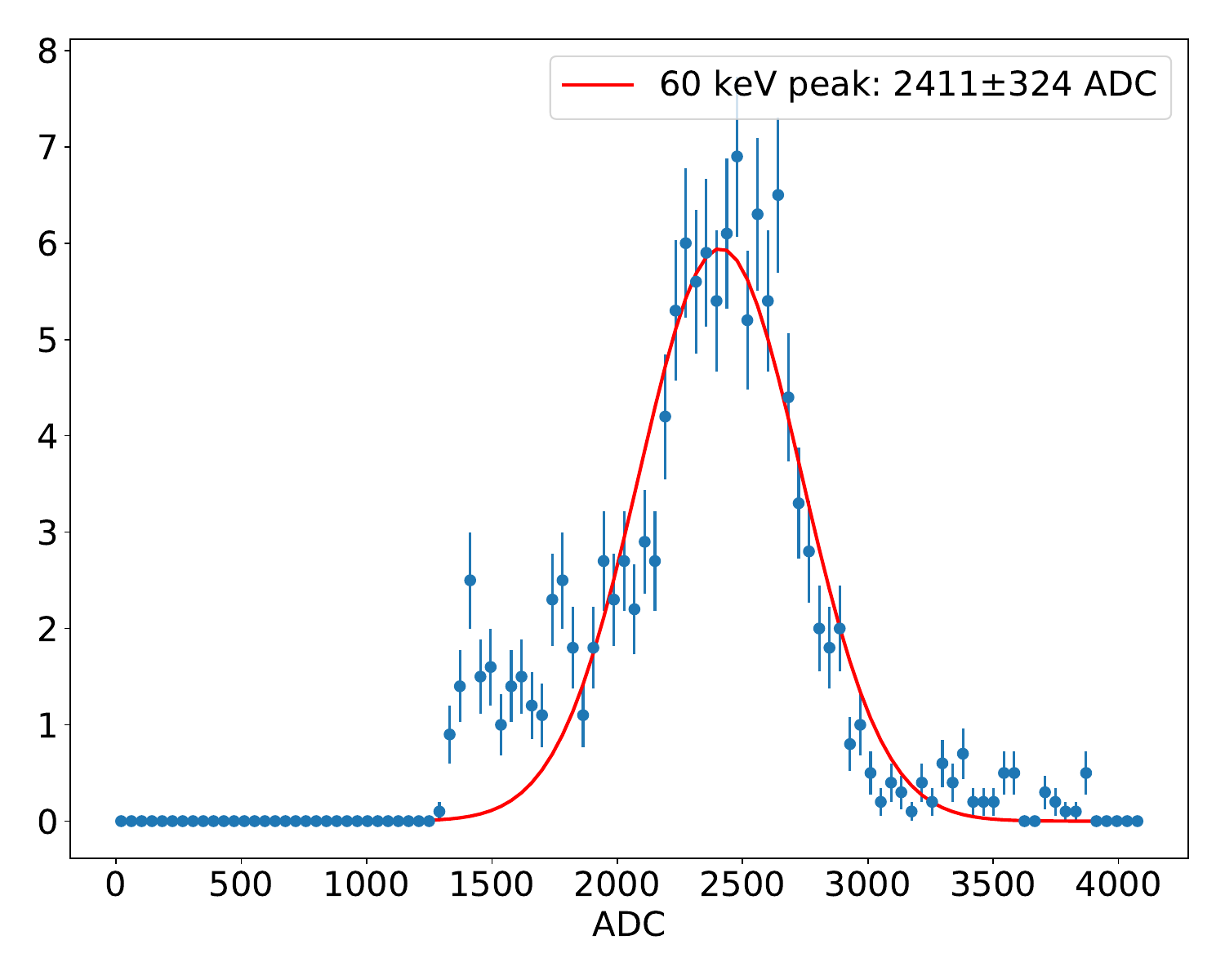}
  \caption{60~keV photo-peak measured with the HG channel at the ESRF facility using an EJ-200 target \citefig{NDA_thesis}.}
  \label{fig:ESRF_photopeak}
\end{figure}

The 64 channels of the module can be scanned with a precise knowledge of the beam position. A 60~keV beam shooting from the module's Zenith direction, whose photo-peak is visible in both HG (see Figure \ref{fig:ESRF_photopeak}) and LG channels (High and Low Gain channels from the CITIROC ASIC \cite{CITIROC_datasheet}), can be used to determine the LY of every channel.

The resulting LY distributions measured for all the channels of EJ-200 and EJ-248M modules are shown in Figure \ref{fig:ESRF_photopeak}. The measured averaged light yields are respectively $1.23\pm0.20$~p.e./keV and $1.37\pm0.32$~p.e./keV. Table \ref{tab:meas_sim_LY} compares the measured light yield values to the simulated ones. Measurements agree with simulations for both types of plastic. For comparison, the light yield of POLAR modules was around 0.3~p.e./keV. The optical efficiency of the polarimeter module has therefore improved considerably, greatly increasing the overall instrument sensitivity.

\begin{table}[H]
\centering
\begin{tabular}{lcc}
\hline
 & LY$_{EJ-200}$ [p.e./keV] & LY$_{EJ-248M}$ [p.e./keV] \\ \hline\hline
Measurement & $1.29\pm0.18$ & $1.40\pm0.26$ \\ \hline
Simulation & $1.27\pm 0.16$ & $1.40\pm 0.15$ \\ \hline
%Simulation & $1.31\pm 0.16$ & $1.43\pm 0.15$ \\ \hline % using 470keV profile
\end{tabular}
\caption{Measured and simulated light yield for EJ-200 and EJ-248M POLAR-2 modules.}
\label{tab:meas_sim_LY}
\end{table}

However, a big spread can be observed for the light yield of the EJ-248M, with two out-sanding groups of scintillator bars. The first group of scintillators has a light yield centered around 1.21~p.e./keV with a standard deviation of 0.31~p.e./keV, while the averaged light yield of the second group is 1.61$\pm$0.19~p.e./keV. The lower light yield group is likely due to a non-uniform optical coupling between the scintillators and the SiPM arrays caused by the repeated assembly-disassembly of the modules during the calibration campaign. Furthermore, light yield of around 1.6~p.e./keV has already been measured for an entire EJ-248M module during preliminary prototypes testing at CERN. Considering this higher light yield for EJ-248M in the case of a good optical contact for the entire module, we notice a deviation from the simulated value.

\begin{figure}[H]
\centering
  \hspace*{-0.5cm}\includegraphics[height=0.45\textwidth]{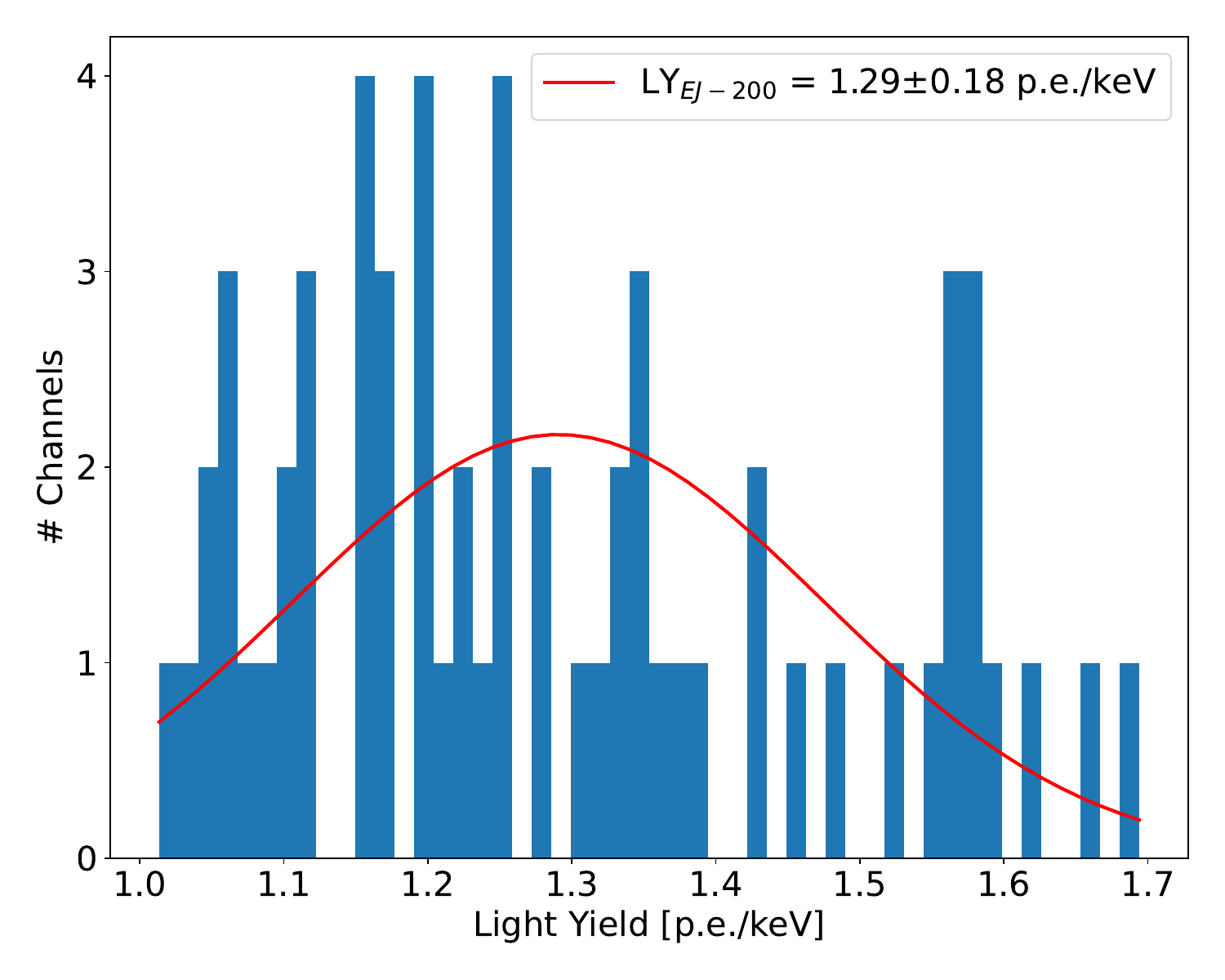}\includegraphics[height=0.45\textwidth]{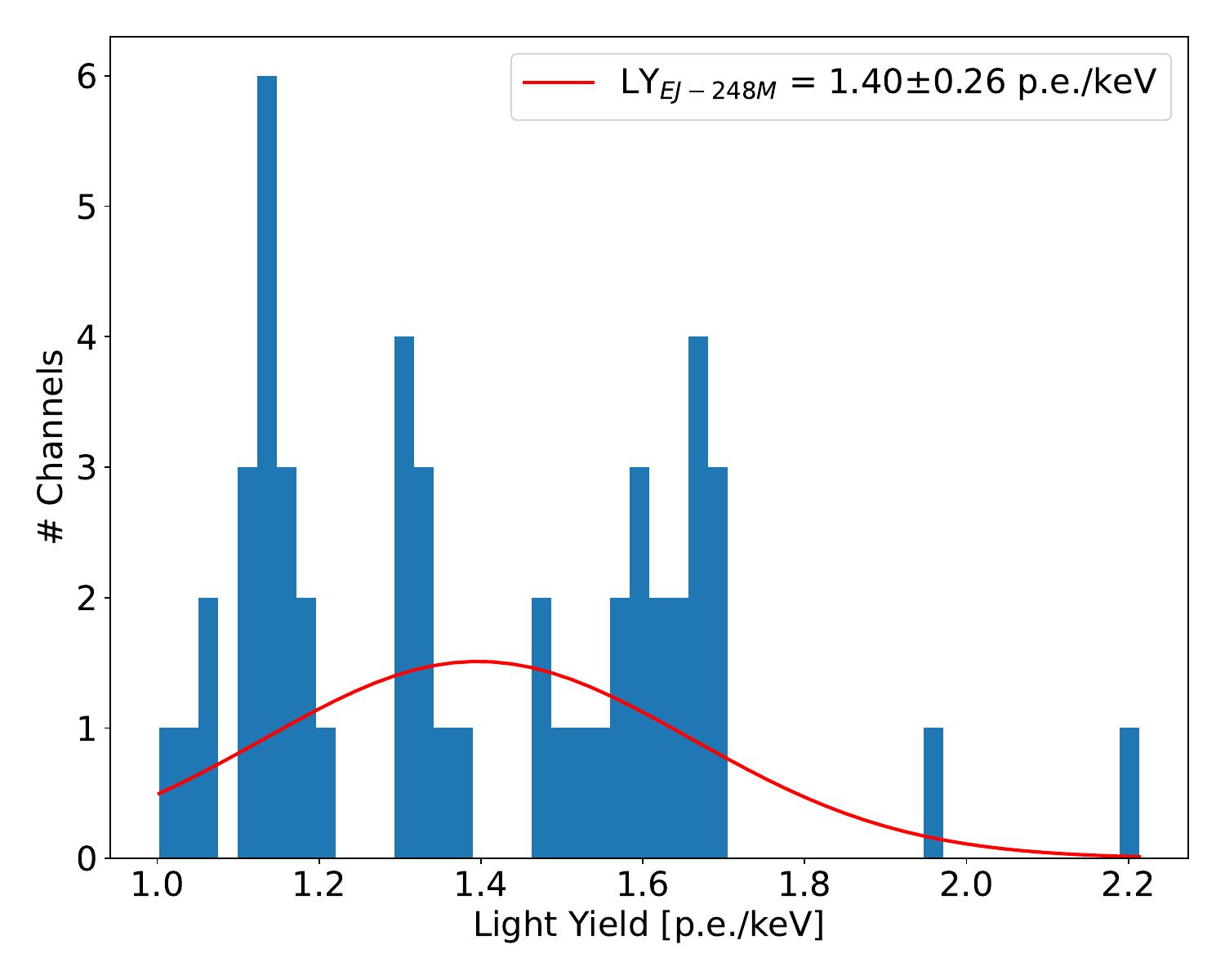}
  \caption{Measured light yield distribution for every channel of an EJ-200 (\textbf{top}) and EJ-248M (\textbf{bottom}) module. Respective light yields of LY$_{EJ-200}=1.29\pm0.18$~p.e./keV and LY$_{EJ-248M}=1.40\pm0.26$~p.e./keV are obtained. It can be noticed in the histograms that only 60 channels are used for the EJ-248M measurement because of malfunctioning channels in the FEE used for this measurement. A big spread can also be observed for the EJ-248M light yield with two groups of channels. This is likely due to bad optical coupling of the group of channels with low light yield, since around 1.6~p.e./keV light yield has been achieved with previous measurements of EJ-248M targets.}
  \label{fig:LY_ESRF}
\end{figure}

This under estimation of the light yield for EJ-248M simulations is likely to be caused by an over estimation of the scintillator roughness for this material. This might be due to the fact that less samples were measured for EJ-248M compared to EJ-200, and that because of samples availability, the EJ-248M bars that were measured on the IOM had already been used in a prototype module. They could therefore have some scratches on the surfaces, leading to a higher measured roughness than for a brand new scintillator.\\

\begin{figure}[H]
\centering
  \hspace*{-0.5cm}\includegraphics[width=0.7\textwidth]{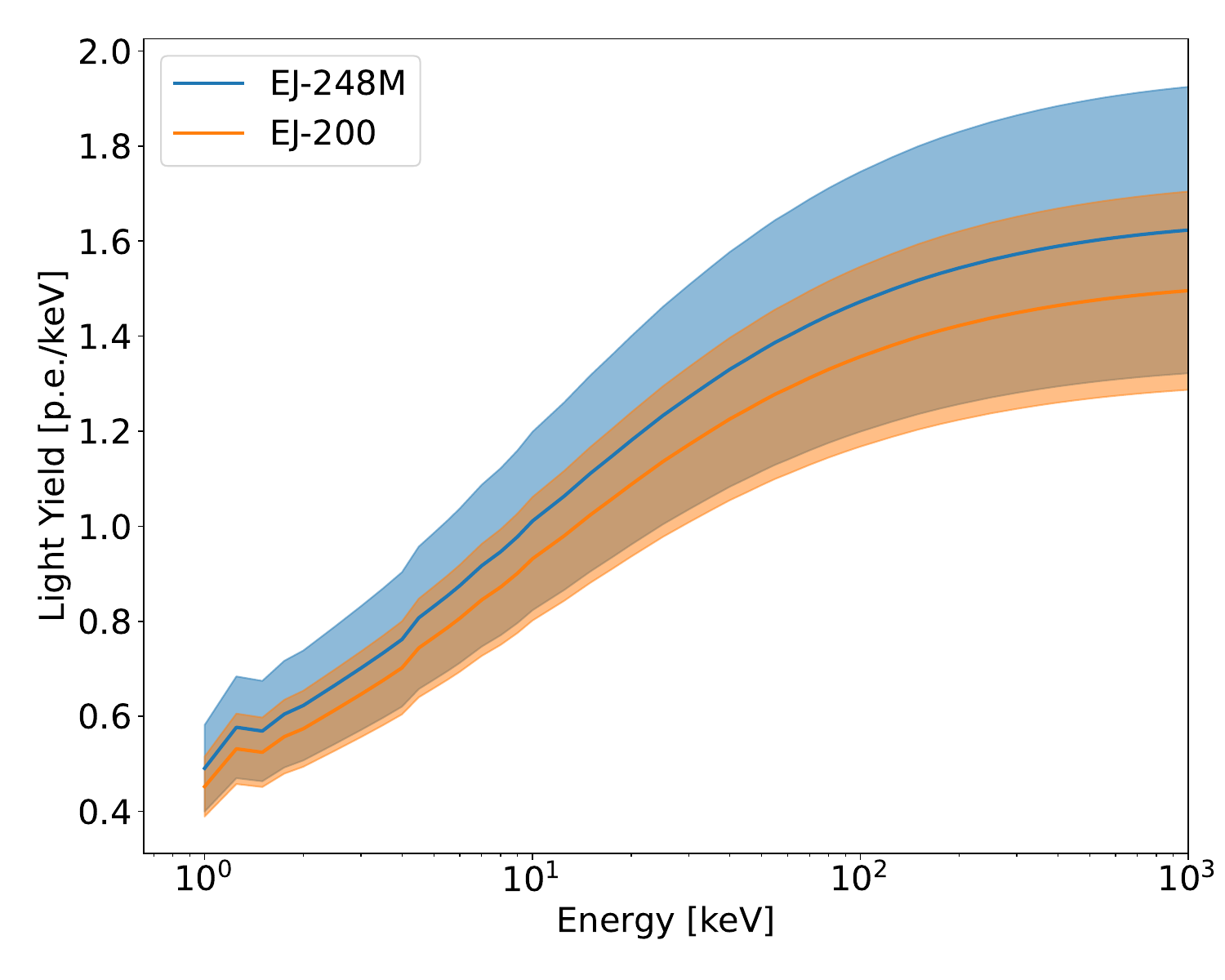}
  \caption{Extrapolated light yield as a function of energy accounting for Birk's effect on the scintillation efficiency for both EJ-200 and EJ-248M materials.}
  \label{fig:LY_birks}
\end{figure}

Finally, the light yield measured at an energy of 60~keV, it has to be corrected for the decrease of scintillation efficiency at low energy using Birk's law. Figure \ref{fig:LY_birks} shows the expected light yield as a function of the incoming photon energy for a Birk's constant of 0.143~mm/MeV \cite{ZHANG201594}.

\subsection{Optical cross talk}

Optical photons can escape from a given scintillator to reach one of the neighbor channels, this is called optical crosstalk and has to properly be measured and simulated in order to fully understand the response of the polarimeter. The optical crosstalk for POLAR was of the order of 10-15\%. Photons going to the wrong readout channel can happen either via propagation through the optical coupling pad, the SiPM resin, the not completely opaque plastic alignment grid, or gaps between the reflective foils used to wrap the major part of the scintillator bars. A feature of the polarimeter module design important for the crosstalk is the fact that the target is divided in four quadrants due to the use of four 16 channels SiPM arrays for the readout. The pitch between channels is therefore 6.2~mm inside a quarter of the module, and 6.5~mm between two neighbour channels that do not belong to the same quarter due to the middle cross spacing between the four SiPM arrays. We expect less crosstalk between the channels separated by the middle cross due to the higher distance separating them.

\begin{figure}[H]
\centering
  \hspace*{-0.5cm}\includegraphics[height=0.45\textwidth]{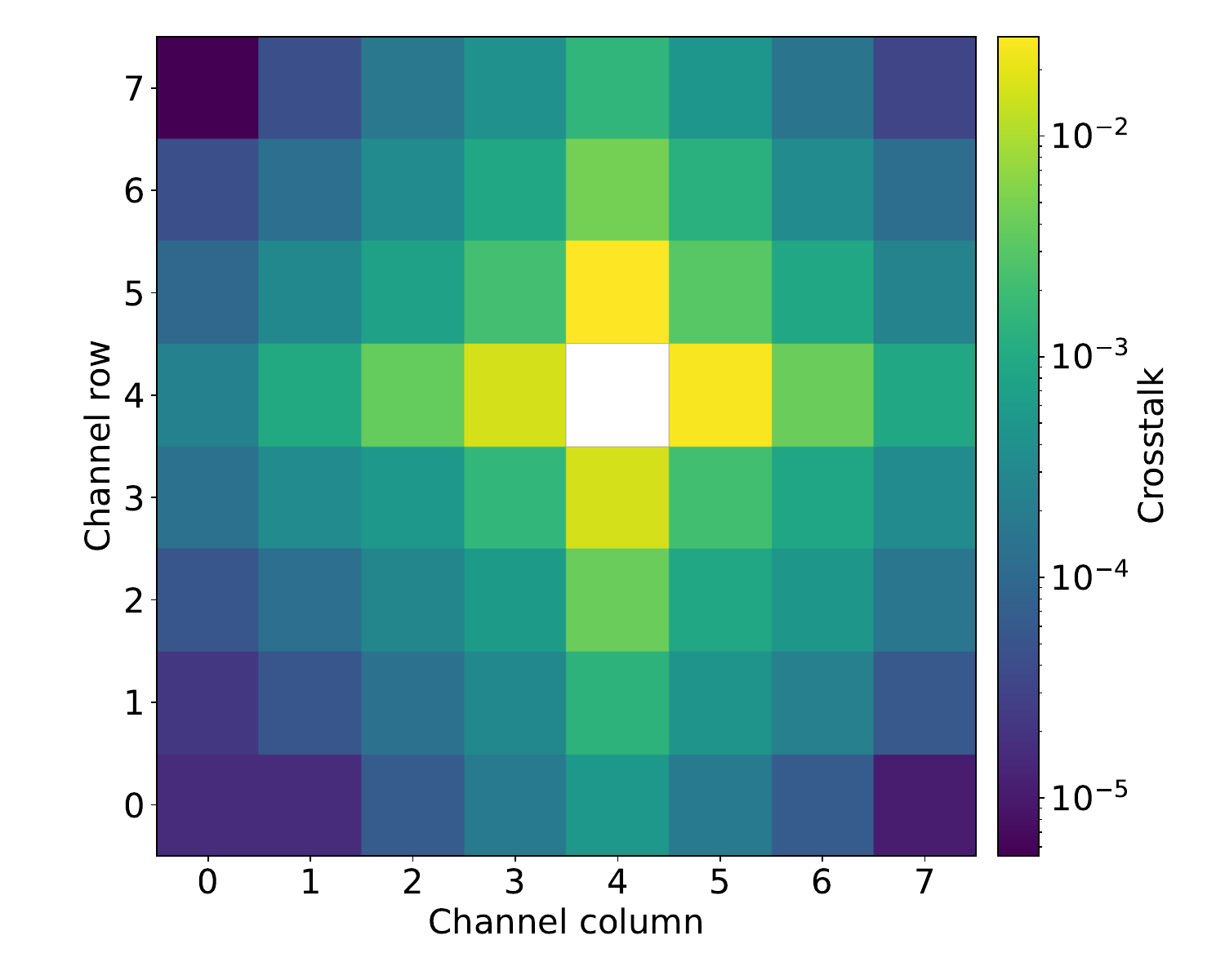}\hspace*{0.1cm}\includegraphics[height=0.45\textwidth]{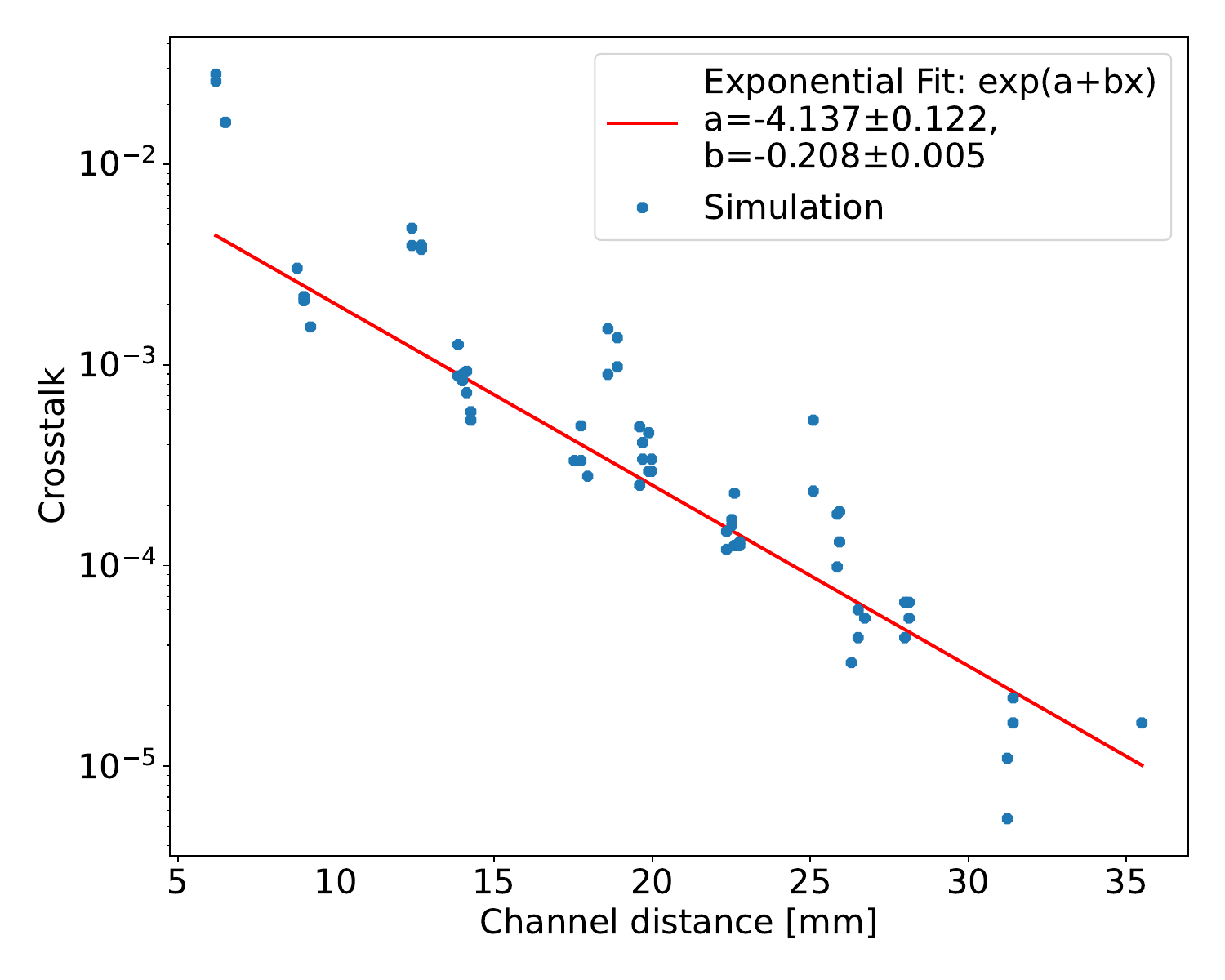}
  \caption{Crosstalk map (\textbf{left}) and crosstalk as a function of the neighbor bar distance (\textbf{right}) for a 150~$\mu$m thick optical coupling pad \citefigadapt{NDA_thesis}.}
  \label{fig:crosstalk_map}
\end{figure}

\begin{figure}[H]
\centering
  \hspace*{-1cm}\includegraphics[height=0.45\textwidth]{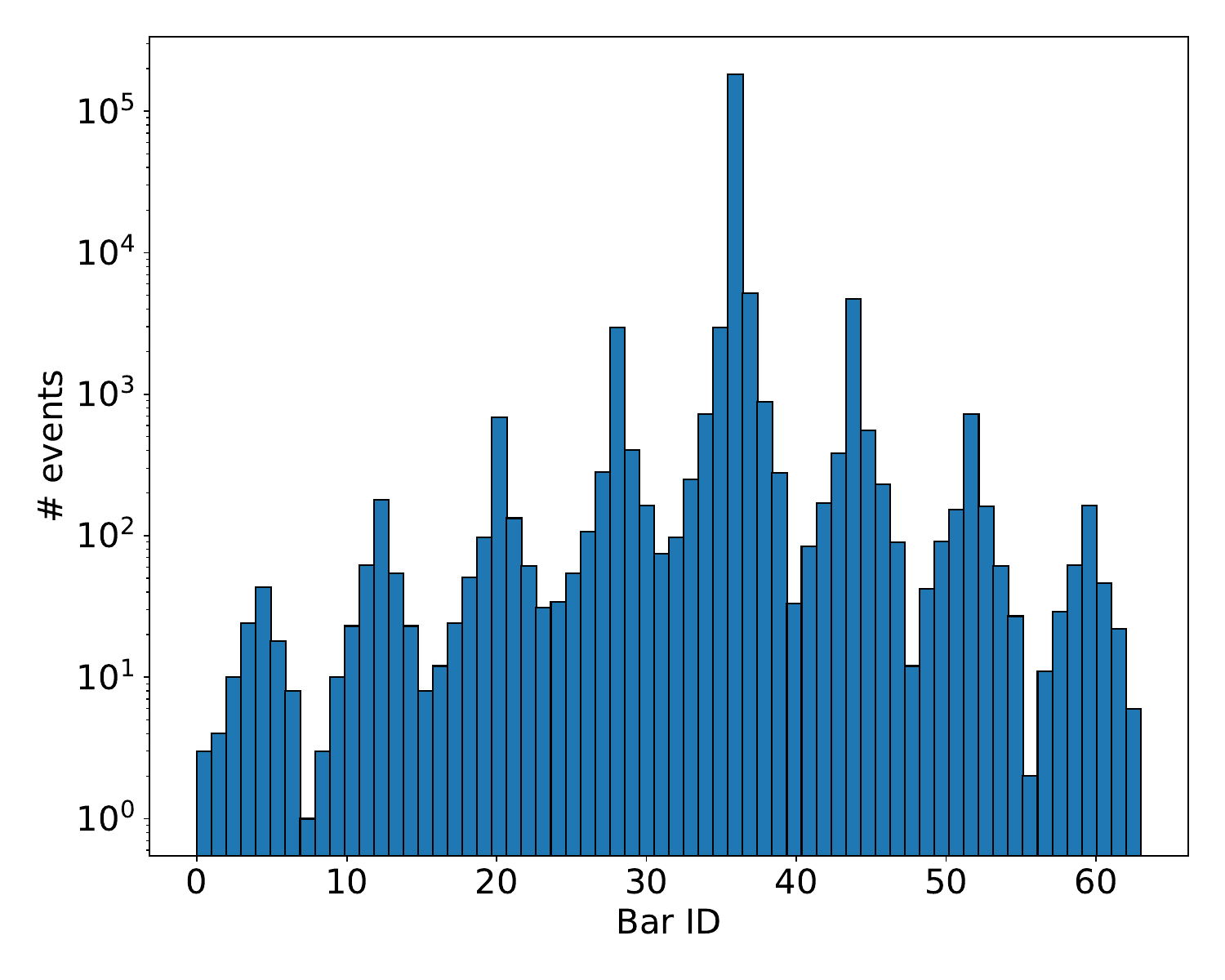}\includegraphics[height=0.45\textwidth]{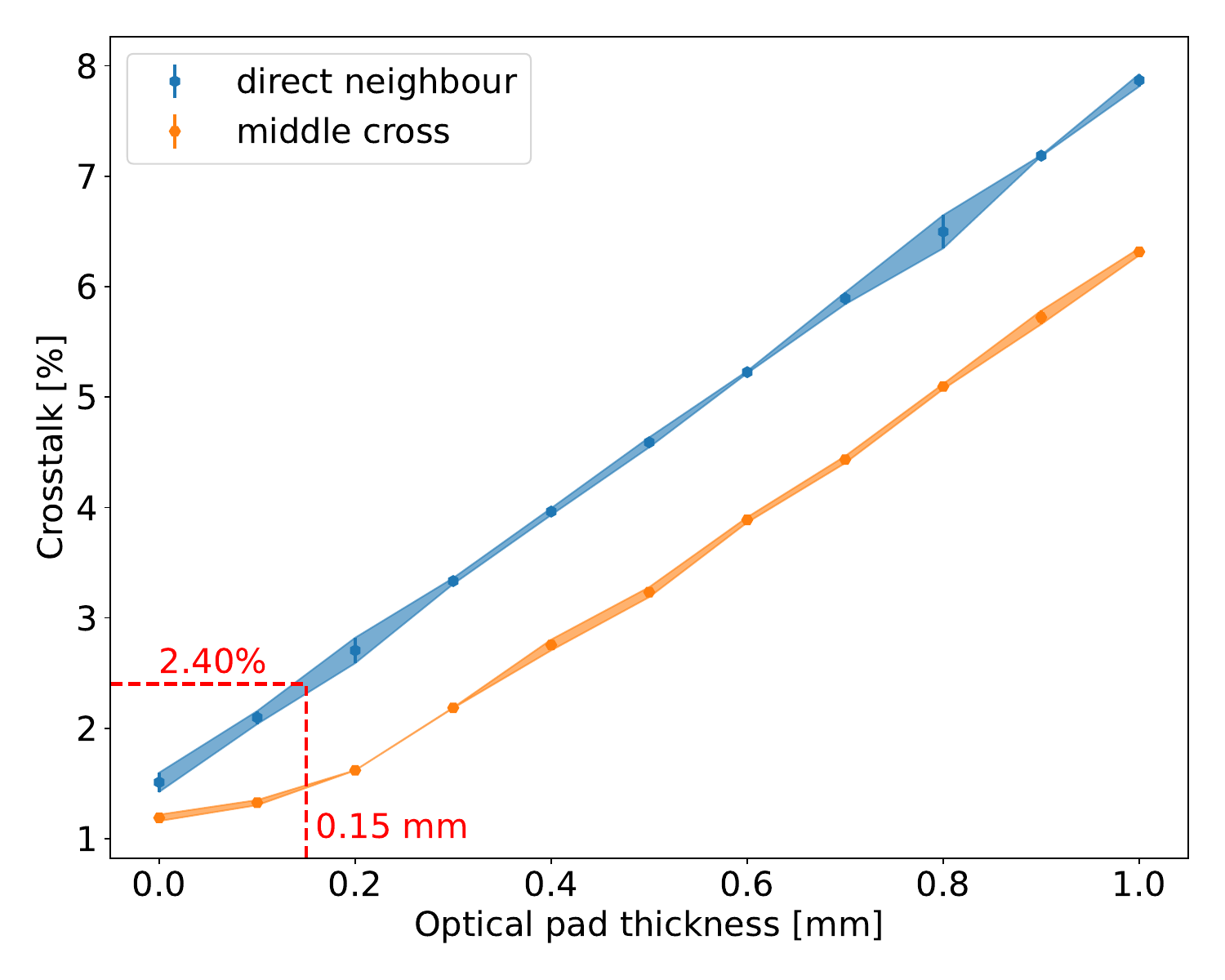}
  \caption{\textbf{Left}: Event distribution in the 8$\times$8 bars array. A clear 8 bar modulation due to the module geometry is observed, with most of the events in the injected channel (\#36). \citefigadapt{NDA_thesis} \textbf{Right:} Optical crosstalk as a function of the optical coupling pad thickness for both 0.2~mm and 0.5~mm pitches (within a SiPM array, and between two arrays where the pitch is bigger to due the middle cross).}
  \label{fig:crosstalk_vs_optpadthickness}
\end{figure}

Optical light is injected in channel \#36 in the simulations, which is located near the center of the module, at the corner of one of the quarters. This way the optical crosstalk to both direct neighbours and neighbours through the middle cross can be studied. An 8$\times$8 map of the simulated crosstalk in each channel of the module for an injection in channel \#36 is plotted in Figure \ref{fig:crosstalk_map}. This figure also shows the crosstalk extracted versus the physical distance to the primary channel, fitted with an exponential. As expected, slightly lower crosstalk is observed in the map to the neighbour through the middle cross than to the direct neighbour. The map is shown as a one-dimensional plot in Figure \ref{fig:crosstalk_vs_optpadthickness}, where the number of optical photons reaching the SiPM is plotted as a function of the channel number. A periodical behavior is observed with a peak every 8 channels due to the geometry of the module, with the peaks going lower and lower as we go away from channel \#36. The crosstalk was simulated here for a 150~$\mu$m thick optical pad, which is used for coupling the scintillator target with the SiPM arrays in the final design. As the crosstalk is dependent on the thickness of the optical pad, the simulations were repeated as a function of this thickness in order to characterize the increase of crosstalk for both direct and "middle cross" neighbour. The result is plotted in Figure \ref{fig:crosstalk_vs_optpadthickness}, with a crosstalk of 2.40\% for a 150~$\mu$m thick pad.

%root output_xtalk04.root >> HitsCol->Draw("BarID>>h2(64,0,63)")

The optical crosstalk can be estimated with a simple calculation to ensure that the simulated crosstalk is reasonable. As the reflecting foil are almost perfectly reflecting all the photons (see Section \ref{subsec:ESR_characterization}), the main crosstalk contributions come from the optical pad and the alignment grid. The contribution of the alignment grid can be calculated by multiplying the fraction of scintillator height that it covers, 3~mm/125~mm=2.4\%, by the grid transmittance of 60.9\% for a 0.2~mm thickness (see Section \ref{subsec:plastic_grid}), which leads to a crosstalk of 1.46\%. Doing the same calculation for a "middle cross" neighbour, where the grid is 0.5~mm thick and the transmittance is 47.1\%, one obtains a crosstalk of 1.13\%. Both crosstalk contributions from the grid are compatible with the value obtained for a null optical pad thickness in Figure \ref{fig:crosstalk_vs_optpadthickness}, showing the great accuracy of the optical simulations.

\begin{figure}[H]
\centering
  \hspace*{-1cm}\includegraphics[height=0.45\textwidth]{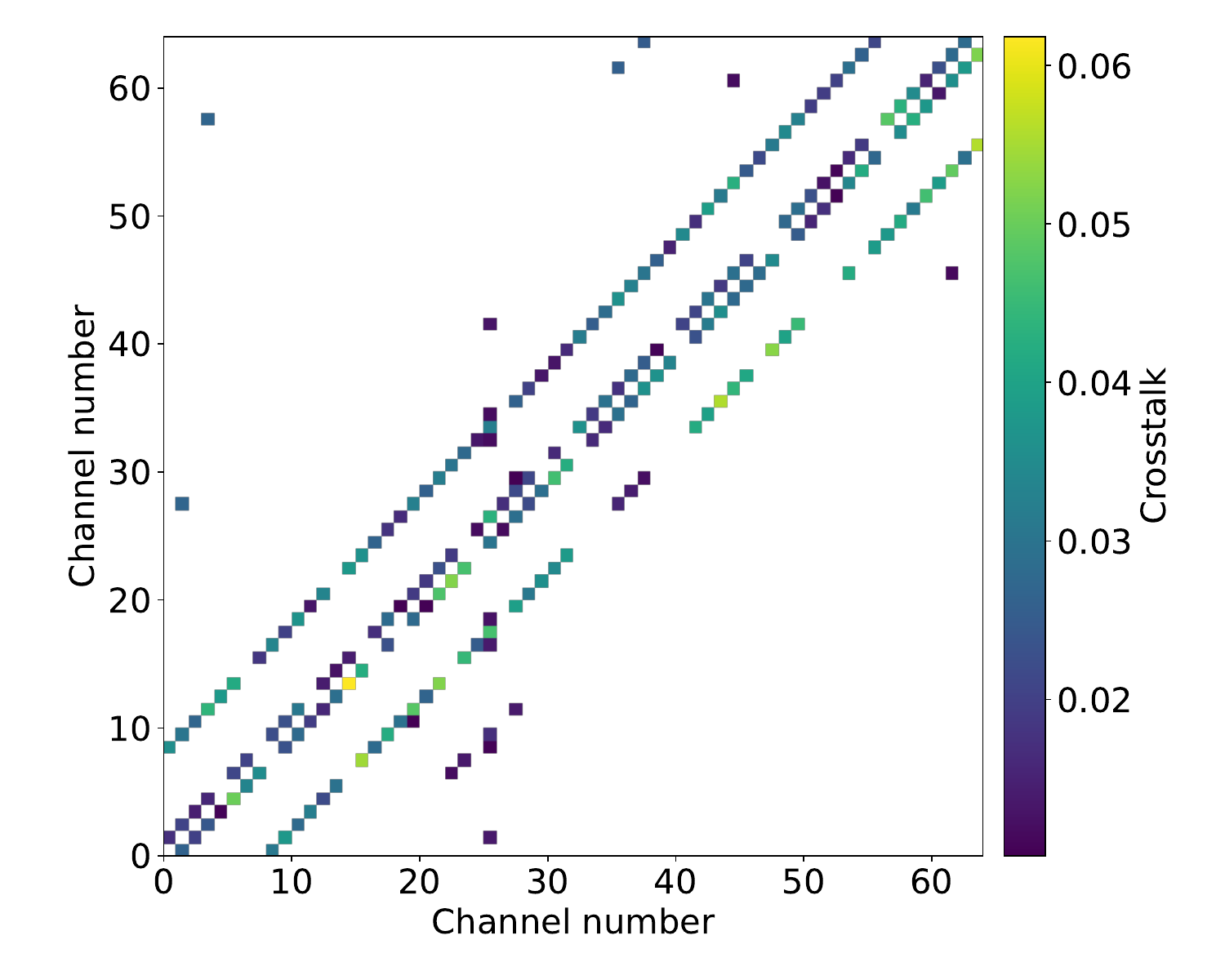}\hspace*{0.1cm}\includegraphics[height=0.45\textwidth]{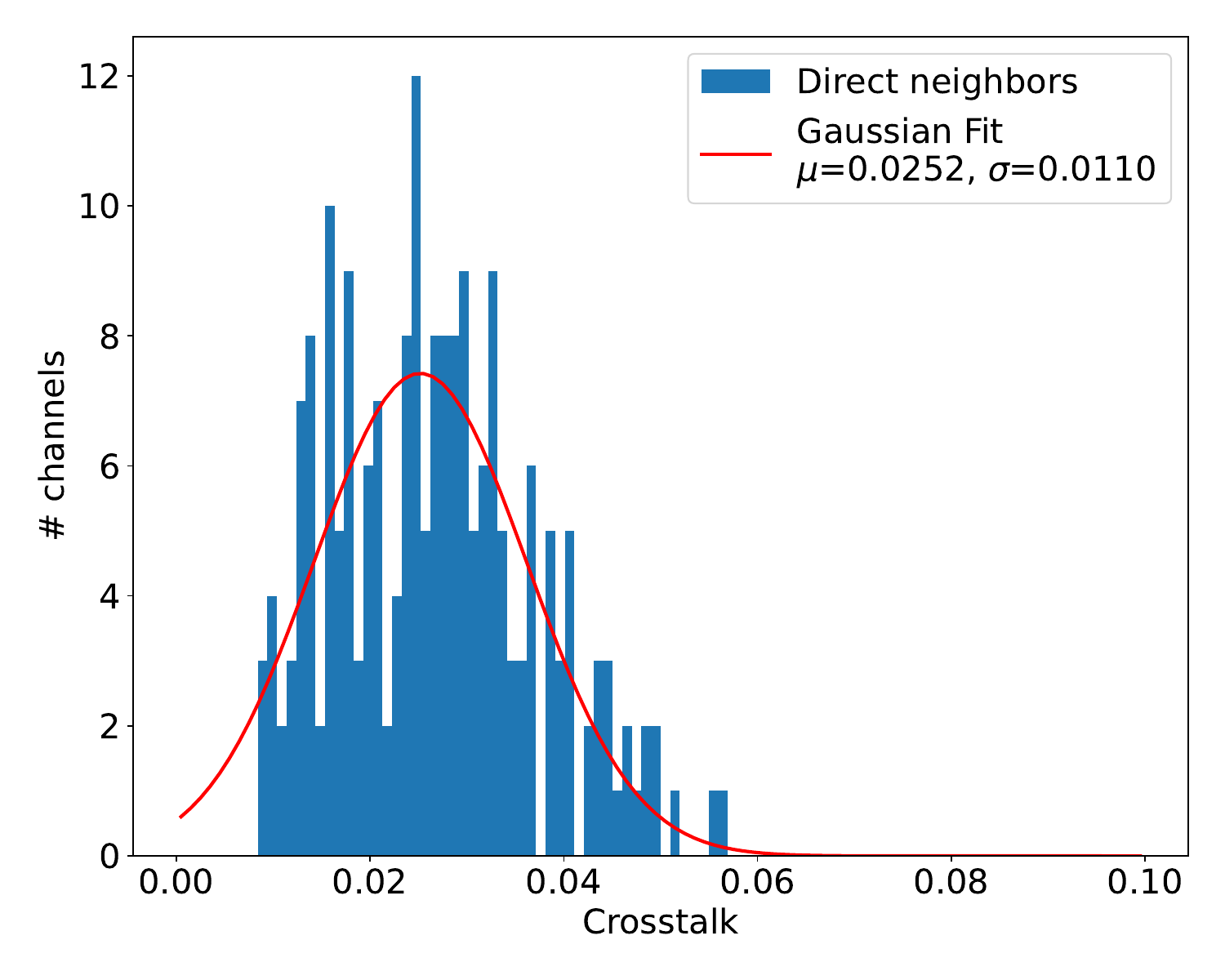}
  \caption{\textbf{Left:} Measured 64$\times$64 crosstalk map showing for each channel the measured crosstalk to the other 63 channels. \textbf{Right:} Distribution of the measured crosstalk extracted from the map for the first neighbours of every of the 64 channels.}
  \label{fig:measured_crosstalk}
\end{figure}

The optical crosstalk for every channel is measured by scanning the module with a 60~keV beam. The measured crosstalk is shown in Figure \ref{fig:measured_crosstalk} for each channel versus each other 63 channels in the form of a 64$\times$64 map. The central diagonal corresponds to the crosstalk in one direction (row), while the two diagonals on each side shifted by 8 channels correspond to the neighbour in the other direction (column) as well as the diagonal neighbour (that do not share an edge but only a corner with the main channel). The other two diagonals on the outside, shifted by 16 channels are crosstalk to the second-order neighbour in the column direction. The crosstalk to the second neighbour in the row direction can be seen around the main diagonal.\\

The crosstalk values to the first neighbour are also filled in a histogram plotted in Figure \ref{fig:measured_crosstalk}. A great improvement compared to POLAR can be observed. As mentioned earlier, the optical crosstalk in POLAR was about 10 to 15\% because of the thick optical pad (1~mm versus 150~$\mu$m for POLAR-2) and of the thick MA-PMT entrance window. An average crosstalk of $2.52\pm 1.10$\% is measured from this histogram, compatible with the 2.40\% simulated value reported in Figure \ref{fig:crosstalk_vs_optpadthickness}.

%\section{Polarimeter Target Calibration}

\newpage
\section{Summary and Outlook}

The role of the POLAR-2 polarimeter modules is to convert the energy deposited by the incoming $\gamma$-rays into optical photons, which will later be collected by light sensors. The conversion of the deposited energy into optical light has to be done with the highest possible efficiency. The polarimeter module is therefore not only a $\gamma$-ray detector, but it is an optical system whose optical components have to be optimized, characterized, and simulated in order to reach the best instrumental performances.\\

The POLAR-2 polarimeter module consists of a target of 8$\times$8 elongated plastic scintillators, individually wrapped in highly reflective foils. The individual wrapped bars are held together with a specially developed mechanical grid, and dampers are placed at the top of the target for vibration dampening. Once inserted into its carbon fiber socket, the target is coupled to the SiPM arrays and their front-end electronics thanks to a thin and soft optical pad.\\

The reflectance and transmittance of several reflective foils have been characterized, as well as the transmissivity of the optical pad and of the mechanical grid used to align the scintillators. Two scintillating materials were investigated for the module: EJ-248M, previously used in POLAR, and EJ-200, which has a 9\% higher scintillation efficiency. The surfaces of several scintillators of both types were scanned with an interference optical microscope to characterize the surface quality of both plastics.\\

Based on the characterization of the different optical components composing the module, a complete optical simulation of a polarimeter module was implemented in Geant4. Among other things, the surface quality of both plastic scintillator candidates was included in the simulation using the $\sigma_\alpha$ roughness parameter in Geant4. Simulations show that although the EJ-200 material has a higher scintillation efficiency, a higher overall optical efficiency is to be expected from EJ-248M, whose surface roughness is lower. Indeed, because the EJ-248M material is harder than the EJ-200, it has a smoother surface and therefore less light is lost at the interfaces, counterbalancing the difference in scintillation efficiency. This was confirmed through lab measurements by measuring the light yield of a module based on EJ-248M and comparing it to that measured for an EJ-200 target. Defined as the number of detected optical photons per amount of deposited energy, the light yield is a good figure of merit to assess the optical efficiency of the module. As expected from simulations, a better light yield was measured with EJ-248M, which was selected as the scintillating material for the final design of the polarimeter. Finally, the optical crosstalk, corresponding to the fraction of optical light escaping from a given channel to one of its neighbour channels, was also simulated. The crosstalk measured in the lab matched the simulated values, with an improvement of almost an order of magnitude compared to POLAR.\\

Finally, it should be noted that the optical simulations developed for POLAR-2 are also useful for other similar instruments. Simulation work for determining the optimal design configuration was conducted for another GRB-dedicated polarimeter, the LargE Area Burst Polarimeter (LEAP) experiment \cite{LEAP_AAS2020, LEAP_proto_SPIE, LEAP_SPIE21}. The simulation work performed for LEAP is described in \cite{NDA_thesis}.

\newpage
\section*{Ackowledgements}

We thank Prof. Marc Jobin from the Micro-Nanotechnology group of the School of Landscape, Engineering and Architecture of Geneva as well as the Laboratory of Advanced Technology (LTA), Geneva, for their help with the interferometric optical measurements of scintillator surface quality. We thank Thomas Schneider from the CERN Thin Film \& Glass service for his support with the optical characterization of reflective foils. We thank Sébastien Clément from the CERN Polymer lab for his technical help on the alignment grid manufacturing. %We are also very thankful to Coralie Husi for her help with the mechanical assembly of the POLAR-2 prototype modules.

We gratefully acknowledge the Swiss Space Office of the State Secretariat for Education, Research and Innovation (ESA PRODEX Programme) which supported the development and production of the POLAR-2 detector. M.K. and N.D.A. acknowledge the support of the Swiss National Science Foundation. National Centre for Nuclear Research acknowledges support from Polish National Science Center under the grant UMO-2018/30/M/ST9/00757. %We gratefully acknowledge the support from the National Natural Science Foundation of China (Grant No. 11961141013, 11503028), the Xie Jialin Foundation of the Institute of High Energy Physics, Chinese Academy of Sciences (Grant No. 2019IHEPZZBS111), the Joint Research Fund in Astronomy under the cooperative agreement between the National Natural Science Foundation of China and the Chinese Academy of Sciences (Grant No. U1631242), the National Basic Research Program (973 Program) of China (Grant No. 2014CB845800), the Strategic Priority Research Program of the Chinese Academy of Sciences (Grant No. XDB23040400), and the Youth Innovation Promotion Association of Chinese Academy of Sciences.

\normalem
\printbibliography

\appendix
\section{Detailed optical characterization of reflective foils for scintillator wrapping}
\label{sec:appendix_ESR}

\begin{figure}[H]
\centering
 \hspace*{-0.5cm}\includegraphics[height=.45\textwidth]{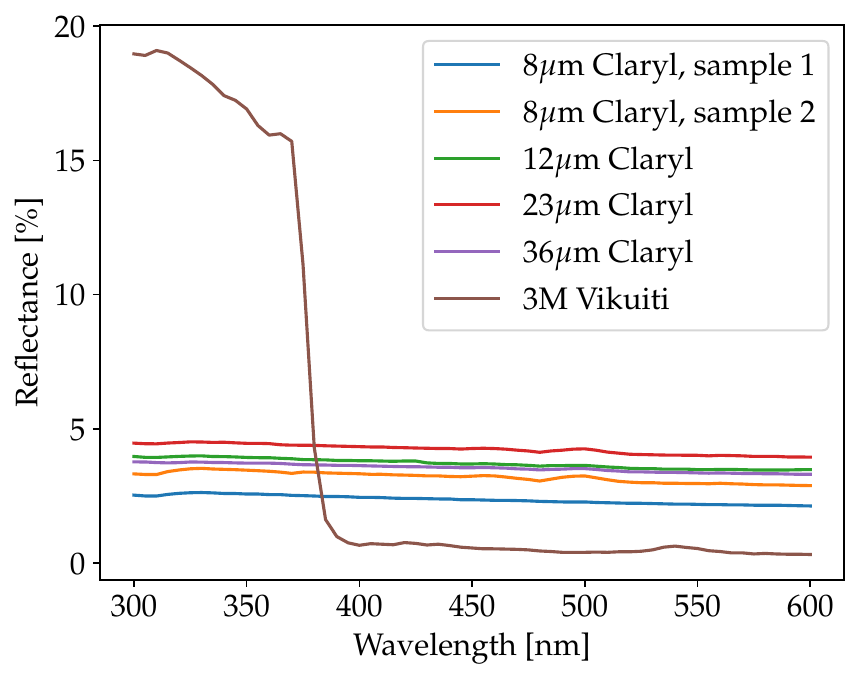}\includegraphics[height=.45\textwidth]{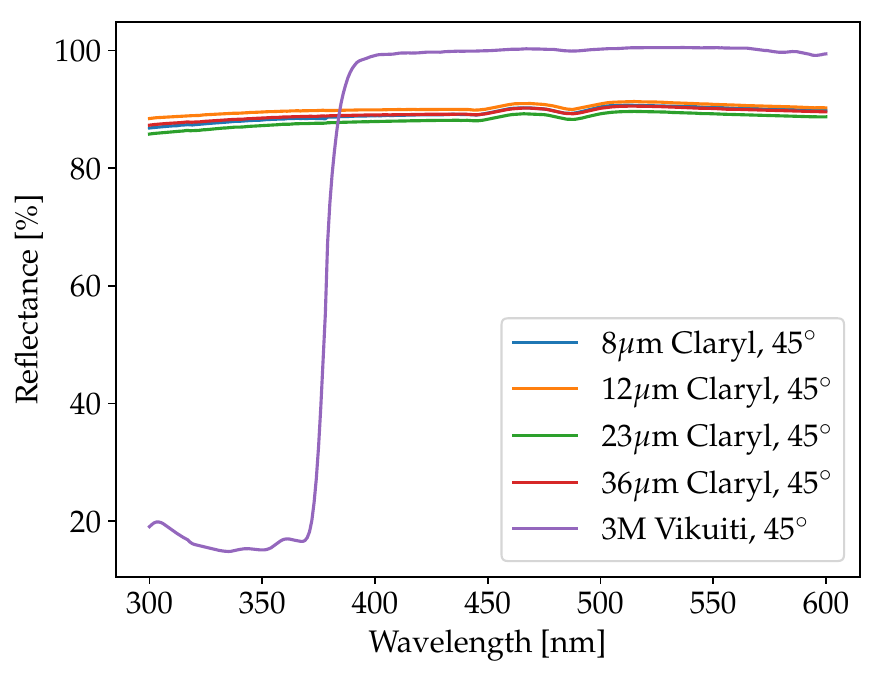}
 \caption{\textbf{Left:} Measured diffuse reflectance spectrum for Vikuiti and Claryl samples with the integration sphere. \textbf{Right:} Measured specular reflectance for Vikuiti and Claryl samples with the URA for a 45$^\circ$ incident angle. \citefig{NDA_thesis}}
 \label{fig:diff_refl_spec_refl}
\end{figure}

Here are provided relevant additional results from the reflective foils characterization discussed in Section \ref{subsec:ESR_characterization}. The diffuse and specular components of the reflectance of the different reflective foils are first provided in Figure \ref{fig:diff_refl_spec_refl}. It can be observed that the Vikuiti is a specular reflector with a diffuse component of the sub-percent level in the visible range, while the Claryl has a bigger diffuse contribution of the order of 3 to 5\%. It can also be noticed that the diffuse component becomes significant for Vikuiti below the 380~nm cutoff, where the specular reflectance contribution to the total reflectance is going down to about 50\%, but this is out of our range of interest.

\begin{figure}[H]
\centering
 \hspace*{-1cm}\includegraphics[height=.45\textwidth]{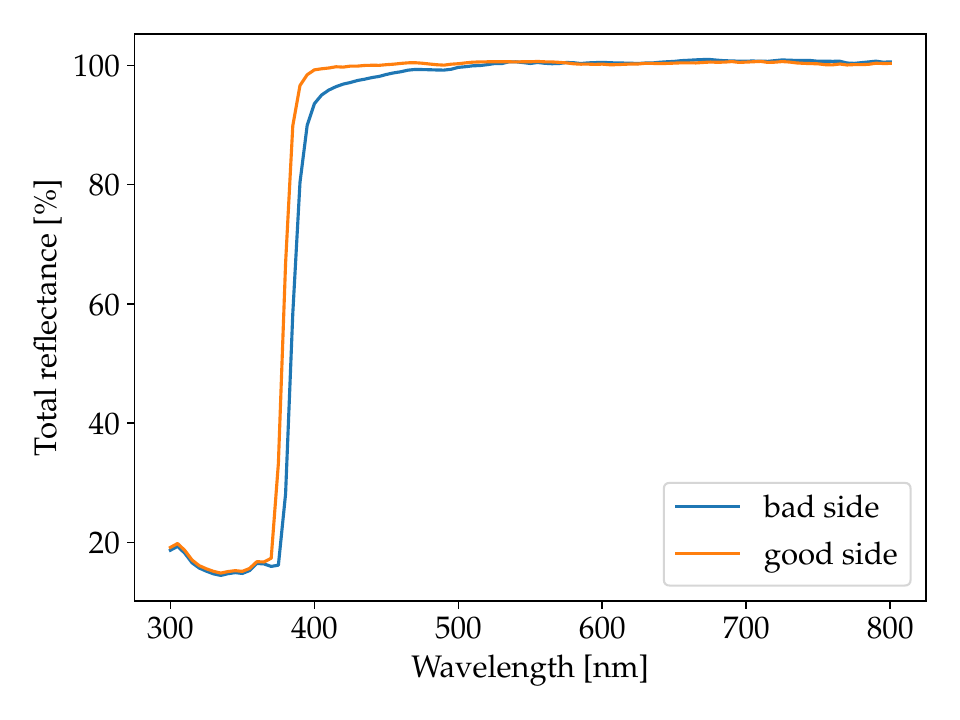}\includegraphics[height=.45\textwidth]{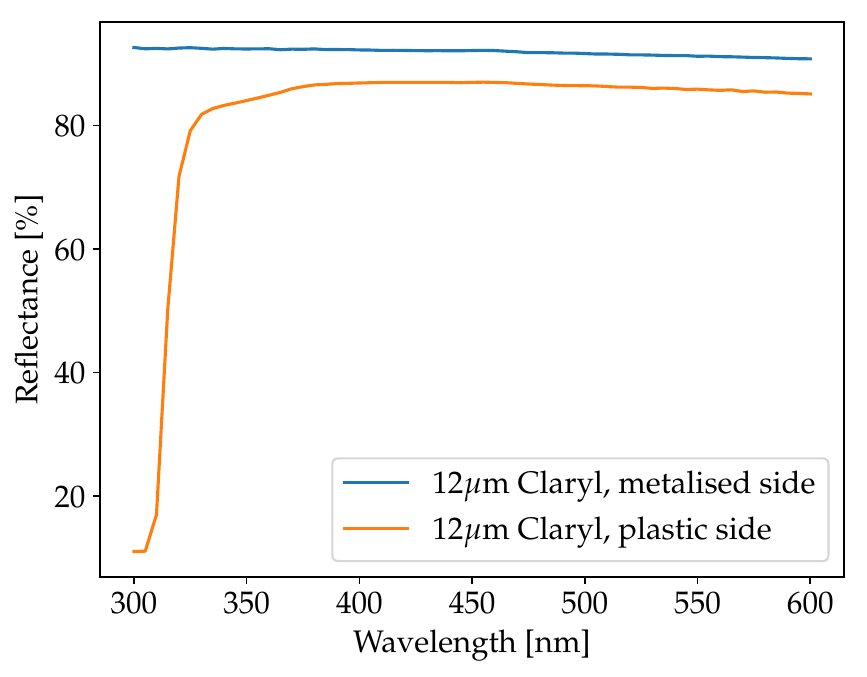}
  \caption[Total reflectance of 3M Vikuiti and Toray Claryl foils measured on both sides.]{\textbf{Left:} Total reflectance of a 3M Vikuiti foil measured on both sides. The reflectance is slightly higher in the 400-500~nm range of interest on one side of the reflective foil. This side should therefore be placed facing the scintillators, to minimize the loss of optical photons. \textbf{Right:} Total reflectance measured for both sides of a Claryl sample. The reflectivity is higher on the metalized side, which is meant to be the reflecting side, while the other side is not as efficient since the photons have to cross the transparent PET layer before being reflected on the Aluminum layer. \citefig{NDA_thesis}}
 \label{fig:comparison_sides_vikuiti_claryl}
\end{figure}

The total reflectance comparison for both sides of a Vikuiti and Claryl foils are plotted in Figure \ref{fig:comparison_sides_vikuiti_claryl}. In the case of the Vikuiti, a small decrease in reflectivity can be observed in the wavelength range of interest. One should therefore ensure that the best-performing side is facing the scintillator when assembling the module. Additionally, the Claryl foils show a lower reflectivity on the non-metalised side. This is to be expected as the foil consists of a PET layer aluminized only on one side, the wrapping orientation is therefore also relevant in this case.

\begin{figure}[H]
\centering
\includegraphics[height=.45\textwidth]{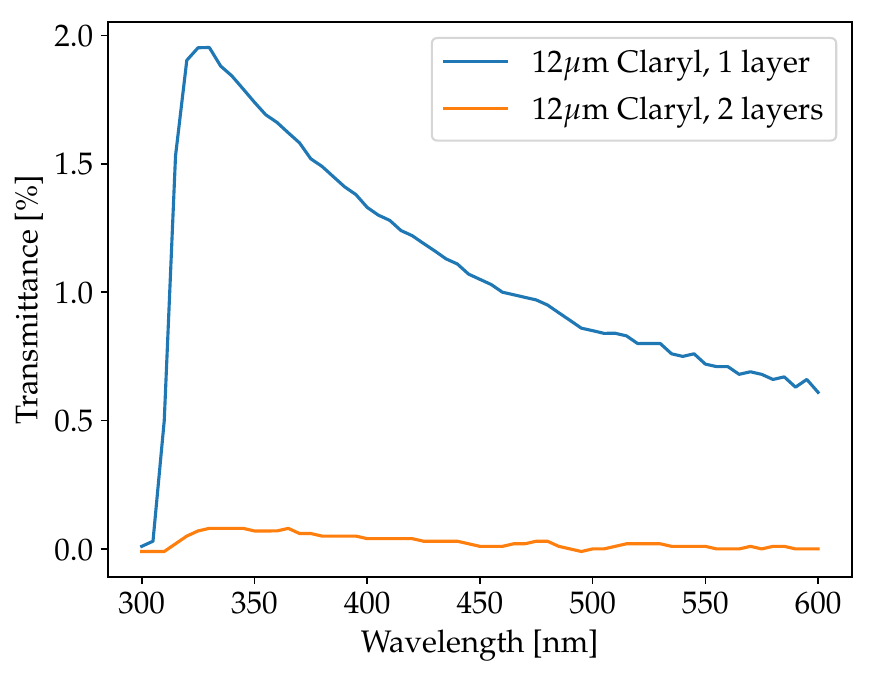}\includegraphics[height=.45\textwidth]{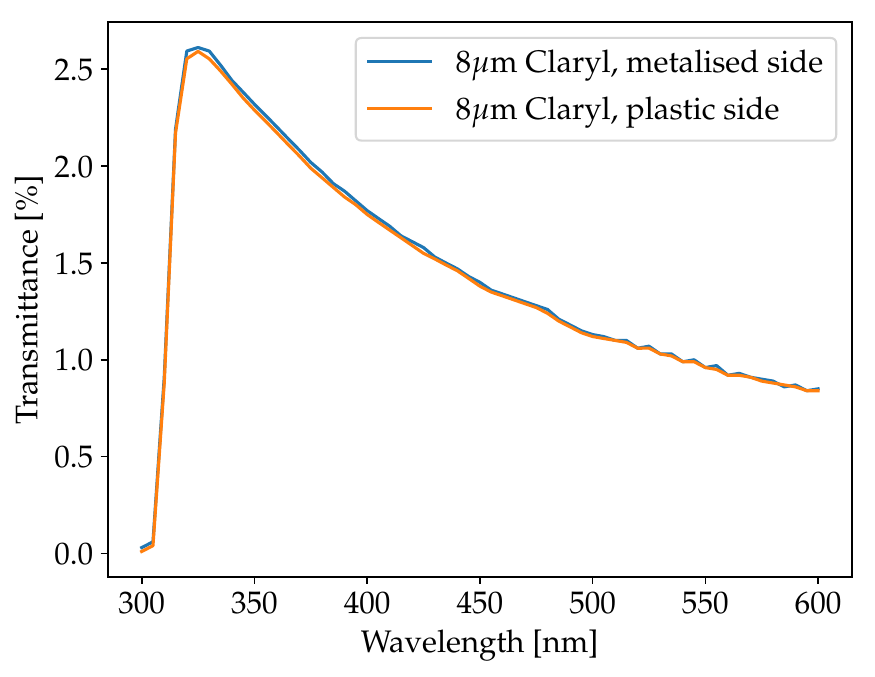}
 \caption{\textbf{Left:} Measured transmittance for one and two layers of 12~$\mu$m thick Claryl. \textbf{Right:} Claryl sample transmittance measured for both sides. The transmittance is not direction-dependent, the spectra are therefore the same. \citefig{NDA_thesis}}
 \label{fig:trans_claryl_comparison}
\end{figure}

The transmittance of two layers of Claryl is compared to that of a single layer in Figure \ref{fig:trans_claryl_comparison}. Adding a second layer reduces the transmittance by almost two orders of magnitude. The transmittance measured on both the plastic and metalized sides of the Claryl is shown in the right plot of the same figure. As one could expect, the transmittance is the same in both directions, while the reflectance is better on the metalized side (see Figure \ref{fig:comparison_sides_vikuiti_claryl}).

\section{Surface Roughness Characterization of Various Plastic Scintillator Materials}
\label{sec:appendix_roughness}

As shown in Sections \ref{subsec:scintillator_jacobian_roughness-meas} and \ref{subsec:roughness_lightoutput}, the surface roughness of plastic scintillators plays an important role in the light loss at the interfaces. It is therefore a crucial parameter for understanding the light collection efficiency of an instrument based on a given plastic scintillator. We report here the measured roughness parameters for other types of plastic scintillators. These scintillator types were not considered for POLAR-2 but could be relevant for other experiments.

\begin{table}[H]
\centering
\begin{tabular}{lccccr}
\hline
Material & $R_a$ [nm] & $R_q$ [nm] & $\sigma_\alpha$ [$^\circ$] & \# samples & Scint. finishing \\ \hline\hline
EJ-200 & $98.1\pm 18.9$ & $123\pm 21$ & $3.45\pm 0.14$ & 10 & d.p. \\ \hline
EJ-248M & $44.0\pm 2.0$ & $62.0\pm 1.0$ & $1.82\pm 0.09$ & 2 & d.p. \\ \hline
EJ-230 & $39.3\pm 3.3$ & $50.0\pm 4.5$ & $1.63\pm 0.11$ & 3 & d.p. \\ \hline
EJ-232 & $48.5\pm 5.5$ & $64.0\pm 6.0$ & $1.91\pm 0.01$ & 2 & d.p. \\ \hline
EJ-232Q0.5\% & $42.0\pm 2.0$ & $56.0\pm 6.0$ & $1.69\pm 0.28$ & 2 & d.p. \\ \hline
EJ-232Q1\% & $47.0\pm 8.0$ & $63.0\pm 7.0$ & $1.92\pm 0.02$ & 2 & d.p. \\ \hline\hline
\textcolor{gray}{EJ-204} & \textcolor{gray}{$13.7\pm 1.7$} & \textcolor{gray}{$18.3\pm 1.2$} & \textcolor{gray}{$1.20\pm 0.03$} & 3 & \textcolor{gray}{a.c.} \\ \hline
\textcolor{gray}{EJ-230} & \textcolor{gray}{$67.0\pm 4.0$} & \textcolor{gray}{$95.0\pm 3.0$} & \textcolor{gray}{$2.76\pm 0.03$} & 2 & \textcolor{gray}{a.c.} \\ \hline
%\textcolor{gray}{EJ-232Q05\%} & \textcolor{gray}{$144\pm 16$} & \textcolor{gray}{$187\pm 21$} & \textcolor{gray}{$4.11\pm0.23$} & 2 & \textcolor{gray}{a.c.} \\ \hline  % problem with the data (interference lines in the corners of the map, the scan range was not big enough)
\end{tabular}
\caption{Measured roughness parameters for various plastic scintillators (d.p.=diamond polished, a.c.="as cast").}
\label{tab:roughness_other_plastics_diamondpolished}
\end{table}

\begin{figure}[H]
\centering
  \includegraphics[width=0.8\textwidth]{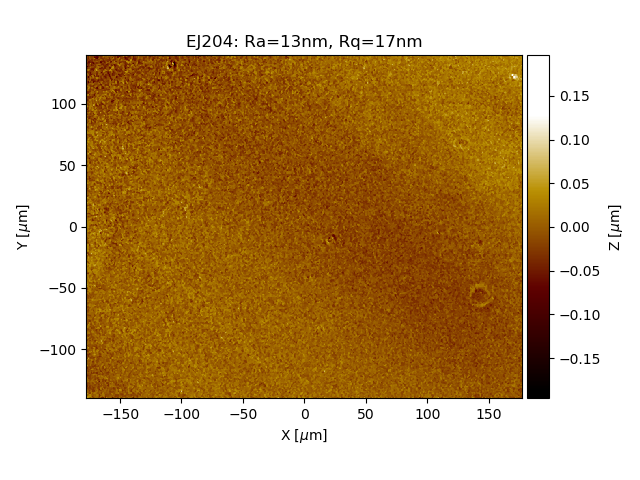}
  \caption{Surface scan measured on a EJ-204 "as-cast" scintillator bar with the IOM.}
  \label{fig:ej204_IOMmap_cast}
\end{figure}

Table \ref{tab:roughness_other_plastics_diamondpolished} gives the measured roughness parameters for various diamond milled plastic scintillators as well as for two "as-cast" scintillators. Despite the latter showing a lower roughness, a great precision on the scintillator dimensions can only be achieve using a diamond milling process. Diamond milling is therefore used for manufacturing the POLAR-2 scintillators. This difference in roughness can clearly be seen between the measured IOM maps of diamond-milled EJ-200 and EJ-248M samples from Figure \ref{fig:ej248m_ej200_IOMmap} and the map measured for "as-cast" EJ-204 shown in Figure \ref{fig:ej204_IOMmap_cast}.

%\begin{figure}[H]
%\centering
%  \includegraphics[width=0.8\textwidth]{Pictures/scintillators/3dmap_EJ200.png}
%  \caption{3D version of the EJ-200 surface scan presented in Figure \ref{fig:ej248m_ej200_IOMmap} \citefig{NDA_thesis}.}
%  \label{fig:ej248m_ej200_IOMmap_3d}
%\end{figure}

\end{document}